\documentclass[aps,showkeys,superscriptaddress,preprint]{revtex4}
\usepackage[latin9]{inputenc}
\usepackage{textcomp}
\usepackage{amsmath}
\usepackage{amssymb}
\usepackage{graphicx}
\usepackage[unicode=true,pdfusetitle,
 bookmarks=true,bookmarksnumbered=false,bookmarksopen=false,
 breaklinks=false,pdfborder={0 0 1},backref=false,colorlinks=false]
 {hyperref}

\makeatletter
\@ifundefined{textcolor}{}
{%
 \definecolor{BLACK}{gray}{0}
 \definecolor{WHITE}{gray}{1}
 \definecolor{RED}{rgb}{1,0,0}
 \definecolor{GREEN}{rgb}{0,1,0}
 \definecolor{BLUE}{rgb}{0,0,1}
 \definecolor{CYAN}{cmyk}{1,0,0,0}
 \definecolor{MAGENTA}{cmyk}{0,1,0,0}
 \definecolor{YELLOW}{cmyk}{0,0,1,0}
}

\usepackage{amsfonts}
\providecommand{\U}[1]{\protect\rule{.1in}{.1in}}
\@ifundefined{textcolor}{}{
\definecolor{BLACK}{gray}{0}
 \definecolor{WHITE}{gray}{1}
 \definecolor{RED}{rgb}{1,0,0}
 \definecolor{GREEN}{rgb}{0,1,0}
 \definecolor{BLUE}{rgb}{0,0,1}
 \definecolor{CYAN}{cmyk}{1,0,0,0}
 \definecolor{MAGENTA}{cmyk}{0,1,0,0}
 \definecolor{YELLOW}{cmyk}{0,0,1,0}
}
\@ifundefined{textcolor}{}{
\definecolor{BLACK}{gray}{0}
 \definecolor{WHITE}{gray}{1}
 \definecolor{RED}{rgb}{1,0,0}
 \definecolor{GREEN}{rgb}{0,1,0}
 \definecolor{BLUE}{rgb}{0,0,1}
 \definecolor{CYAN}{cmyk}{1,0,0,0}
 \definecolor{MAGENTA}{cmyk}{0,1,0,0}
 \definecolor{YELLOW}{cmyk}{0,0,1,0}
}
\@ifundefined{textcolor}{}{
\definecolor{BLACK}{gray}{0}
 \definecolor{WHITE}{gray}{1}
 \definecolor{RED}{rgb}{1,0,0}
 \definecolor{GREEN}{rgb}{0,1,0}
 \definecolor{BLUE}{rgb}{0,0,1}
 \definecolor{CYAN}{cmyk}{1,0,0,0}
 \definecolor{MAGENTA}{cmyk}{0,1,0,0}
 \definecolor{YELLOW}{cmyk}{0,0,1,0}
}
\@ifundefined{textcolor}{}{
\definecolor{BLACK}{gray}{0}
 \definecolor{WHITE}{gray}{1}
 \definecolor{RED}{rgb}{1,0,0}
 \definecolor{GREEN}{rgb}{0,1,0}
 \definecolor{BLUE}{rgb}{0,0,1}
 \definecolor{CYAN}{cmyk}{1,0,0,0}
 \definecolor{MAGENTA}{cmyk}{0,1,0,0}
 \definecolor{YELLOW}{cmyk}{0,0,1,0}
}
\@ifundefined{textcolor}{}{
\definecolor{BLACK}{gray}{0}
 \definecolor{WHITE}{gray}{1}
 \definecolor{RED}{rgb}{1,0,0}
 \definecolor{GREEN}{rgb}{0,1,0}
 \definecolor{BLUE}{rgb}{0,0,1}
 \definecolor{CYAN}{cmyk}{1,0,0,0}
 \definecolor{MAGENTA}{cmyk}{0,1,0,0}
 \definecolor{YELLOW}{cmyk}{0,0,1,0}
}
\@ifundefined{textcolor}{}{
\definecolor{BLACK}{gray}{0}
 \definecolor{WHITE}{gray}{1}
 \definecolor{RED}{rgb}{1,0,0}
 \definecolor{GREEN}{rgb}{0,1,0}
 \definecolor{BLUE}{rgb}{0,0,1}
 \definecolor{CYAN}{cmyk}{1,0,0,0}
 \definecolor{MAGENTA}{cmyk}{0,1,0,0}
 \definecolor{YELLOW}{cmyk}{0,0,1,0}
}
\@ifundefined{textcolor}{}{
\definecolor{BLACK}{gray}{0}
 \definecolor{WHITE}{gray}{1}
 \definecolor{RED}{rgb}{1,0,0}
 \definecolor{GREEN}{rgb}{0,1,0}
 \definecolor{BLUE}{rgb}{0,0,1}
 \definecolor{CYAN}{cmyk}{1,0,0,0}
 \definecolor{MAGENTA}{cmyk}{0,1,0,0}
 \definecolor{YELLOW}{cmyk}{0,0,1,0}
}

\@ifundefined{showcaptionsetup}{}{ \PassOptionsToPackage{caption=false}{subfig}}
\@ifundefined{showcaptionsetup}{}{ \PassOptionsToPackage{caption=false}{subfig}}
\@ifundefined{showcaptionsetup}{}{ \PassOptionsToPackage{caption=false}{subfig}}
\@ifundefined{showcaptionsetup}{}{ \PassOptionsToPackage{caption=false}{subfig}}

\@ifundefined{showcaptionsetup}{}{%
 \PassOptionsToPackage{caption=false}{subfig}}
\usepackage{subfig}
\usepackage{orcidlink}
\usepackage{amsfonts}

\makeatother

\begin{document}
\title{{\Large{}{}Emergence of squeezed coherent states in Kaluza-Klein
cosmology}}
\author{A. S. Lemos \orcidlink{0000-0002-3940-0779}}
\email{adiel@ufersa.edu.br}
\affiliation{Departamento de Ci\^encias Exatas e Tecnologia da Informa\c{c}\~ao, Universidade Federal Rural do Semi-\'Arido, 59515-000 Angicos, Rio Grande do Norte, Brazil}
\affiliation{Departamento de F\'{i}sica, Universidade Federal de Campina Grande, Caixa
Postal 10071, 58429-900 Campina Grande, Para\'{i}ba, Brazil}

\author{A. S. Pereira \orcidlink{0000-0002-5931-6224}}
\email{alberto.pereira@ifpb.edu.br}

\affiliation{Instituto Federal da Para\'{i}ba, 58755-000 Princesa Isabel, Para\'{i}ba, Brazil}
\affiliation{Departamento de F\'{i}sica, Universidade Federal de Campina Grande, Caixa
Postal 10071, 58429-900 Campina Grande, Para\'{i}ba, Brazil}
\author{F. A. Brito \orcidlink{0000-0001-9465-6868}}
\email{fabrito@df.ufcg.edu.br}

\affiliation{Departamento de F\'{i}sica, Universidade Federal de Campina Grande, Caixa
Postal 10071, 58429-900 Campina Grande, Para\'{i}ba, Brazil}
\affiliation{Departamento de F\'{i}sica, Universidade Federal da Para\'{i}ba, Caixa Postal
5008, 58051-970 Jo\~{a}o Pessoa, Para\'{i}ba, Brazil}
\author{Joao R. L. Santos \orcidlink{0000-0002-9688-938X}}
\email{joaorafael@df.ufcg.edu.br}

\affiliation{Departamento de F\'{i}sica, Universidade Federal de Campina Grande, Caixa
Postal 10071, 58429-900 Campina Grande, Para\'{i}ba, Brazil}
\begin{abstract}
In this work, we consider a propagating scalar field on Kaluza-Klein-type
cosmological background. It is shown that this geometrical description
of the Universe resembles - from a Hamiltonian standpoint - a damped
harmonic oscillator with mass and frequency, both time-dependents.
In this scenario, we construct the squeezed coherent states (SCSs)
for the quantized scalar field by employing the invariant operator
method of Lewis-Riesenfeld (non-Hermitian) in a non-unitary approach.
The non-classicality of SCSs has been discussed by examining the quadrature
squeezing properties from the uncertainty principle. Moreover, we
compute the probability density, which allows us to investigate whether
SCSs can be used to seek traces of extra dimensions. We then analyze
the effects of the existence of supplementary space on cosmological
particle production in SCSs by considering different cosmological
eras. 
\end{abstract}
\maketitle

\section{Introduction}

The expansion of the Universe has been extensively studied in the
literature \citep{Peebles2003,Oks2021}, being mainly motivated by
a substantial observational and experimental basis \citep{Riess1998,Perlmutter1999}.
Along with cosmic radiation, Universe expansion is an essential component
of the $\Lambda$CDM cosmological model. Assuming that the Universe
is homogeneous and isotropic its geometrical description is given
by the Friedmann-Lema\^{i}tre-Robertson-Walker (FLRW) cosmological background
\citep{Ellis1999}. Furthermore, from higher-dimensional embedding
theories, cosmological models in a Kaluza-Klein scenario have been
presented as possible extensions to the FLRW model \citep{Leon1988,Wesson1994,Wesson2002,Bellini2003}.

In its turn, the intersection of cosmology and quantum optics has
been the focus of increasing attention. As an example, we can highlight
the problem of cosmological particle creation in a time-dependent
gravitational field \citep{Parker1968,Parker1969,Parker1971}, which,
afterwards, has been addressed by the adoption of the language of
squeezed states (SSs) \citep{Grishchuk1989,Grishchuk1990,Hu1994,Ford2021}.
In Refs. \citep{Grishchuk1989,Grishchuk1990}, it was shown that the
vacuum fluctuations should produce relic gravitons, and, in its turn,
in the course of cosmological expansion would be squeezed into strongly
SSs owing to the interaction with the external gravitational field.
This study has been carried out considering the FLRW cosmological
background and can provide valuable insights into the physical conditions
of the early Universe.

Although the classical and quantum propagation of scalar fields is
well developed in the context of the Minkowski metric \citep{Parker2009,DeWitt1975},
its investigation into several background spacetime can provide important
information on the behavior of quantum systems in curved spacetime
and their possible mathematical framework similarity with another
physical systems \citep{Pedrosa2021}. Moreover, since the gravitational
field universally couples to all fields, it is interesting to investigate
the quantum physical systems in curved spacetime, even in the classical
regime. Therefore, quantum optics formalism has been applied to the
study and construction of SSs of scalar fields produced in several
cosmological scenarios \citep{Pedrosa2021,Bertoni1998,Pedrosa2004,Suresh2004,Suresh2004.2},
in order to deal with quantum effects in cosmology. In Ref. \citep{Alencar2012},
for instance, it was investigated a quantized field propagating on
$D$\textminus dimensional de Sitter spacetime and derived the corresponding
Schr\"{o}dinger states employing the invariant operator method of Lewis-Riesenfeld
\citep{Lewis1969,Dodonov1975,Dodonov2003}.

In this work, we intend to investigate the effects of hidden dimensions
on a quantum system in a flat non-static spacetime background. So,
we consider a quantum scalar field propagating on a $5D$ Kaluza--Klein
cosmological background. Here, we assume that the gravitational field,
produced by the curvature of spacetime, is a classical background
where a massless quantum scalar field propagates. For our purposes,
this problem may be reduced to a damped harmonic oscillator with both
mass and frequency time-dependent. Furthermore, we employ the invariant
operator method to construct time-dependent SCSs by applying a non-unitary
approach \citep{Pereira2021,Pereira2023}. These states satisfy the
Schr\"{o}dinger equation for the Hamiltonian of the system associated
with the scalar field. In its turn, in principle, the inflaton field
can be found in coherent states (CSs) and SSs beyond the inflationary
era \citep{Suresh2004,Suresh2004.2} -- at least when inflation ends,
and the scalar field begins to oscillate \citep{Baumann2009}. Thus,
here we intend to explore the possibility of finding the inflaton-like
field in SCSs during the radiation- and matter-dominated eras, aiming
to seek for traces of extra dimensions. For this purpose, we evaluate
the transition probability from a system initially prepared in a Fock
state to the SCSs of a time-dependent damped harmonic oscillator.
The probability density has been calculated and investigates the effects
due to the extra dimension on this quantity during different cosmic
eras. Finally, we compute the time-dependent number of particles produced
in SCSs.We shown that the time-oscillation on the number of created
particles arises owing to the existence of an extra spatial dimension.

The structure of this paper is divided in the following nutshell:
section \ref{sec2} introduces a massless scalar field propagating
in the Kaluza-Klein class of a one-parameter $5D$ cosmological model.
The quantum description of this system and the discussion about the
integral of motion method is the subject of section \ref{sec3}. In
its turn, in section \ref{sec4}, we construct the SCSs and study
the non-classicality of the given system. Besides, we also have presented
plots of the probability density and discussed its results. In section
\ref{sec5}, we study the particle production in the SCSs through
the evolution in time of the number operator. Section \ref{sec6}
contains our final remarks and address insights for future works.

\section{Propagating scalar field in Kaluza-Klein cosmologies\label{sec2}}

This section seeks to reformulate the quantum field propagating issue
in a $5D$ cosmological model such as a time-dependent oscillator.
Let us start with the Kaluza-Klein cosmological $5D$ line element
given by \citep{Wesson2002} 
\begin{equation}
ds^{2}=\ell^{2}dt^{2}-A\left(t,\ell\right)\left(dx^{2}+dy^{2}+dz^{2}\right)-B\left(t\right)d\ell^{2},\label{metric}
\end{equation}
where $\ell$ is the extra-coordinate, and 
\begin{equation}
A\left(t,\ell\right)=t^{2/\alpha}\ell^{2/\left(1-\alpha\right)},\text{ and}\text{ \ }B\left(t\right)=\alpha^{2}\left(1-\alpha\right)^{-2}t^{2}.\label{2}
\end{equation}
In the complementary $4D$ models, the nature of matter is defined
by the $\alpha$ parameter and encompasses both inflationary and non-inflationary
scenarios \citep{Wesson2002}. We point out that for $\alpha=3/2$,
the Eq. (\ref{metric}) describes the late Universe (matter era),
while for $\alpha=2$ will describe the early Universe (radiation
era). Besides, one can consider $\alpha\in\left(0,1\right)$ aiming
to eventually describe inflationary scenarios.

The Einstein-Hilbert action, which governs the theory of five-dimensional
gravity coupled to a scalar field in the $5D$ vacuum Universe can
be written as 
\begin{equation}
S=\int d^{5}x\sqrt{\left|g\right|}\left[\frac{\hat{R}}{16\pi\hat{G}}+\mathcal{L}\left(\Phi,\Phi_{,M}\right)\delta\left(\ell-L\right)\right],\label{action}
\end{equation}
where $g$ is the determinant of the metric tensor, $\hat{R}$ is
the $5D$ Ricci scalar, and $\hat{G}$ is the five-dimensional gravitational
constant. The Dirac delta function is accountable by taking a ``foliation"
(choice of a hypersurface) $\ell=L$ on the metric once we are interested
in studying how is the effective $4D$ dynamics of $\Phi$. On its
turn, in this background, hypersurfaces with $\ell=\mathrm{constant}$
are geometrically identical to spatially-flat $4D$ FLRW cosmological
models. Besides, the building of hypersurfaces $4D$, \textit{via}
foliation, ensures that the equation of state of the induced matter
correctly describes the late (dust) and early (radiation) Universe
\cite{Wesson2002}. In this scenario, we restrict our attention to
a minimally coupled real massless scalar field $\Phi\left(\vec{X},t\right)$,
with $\vec{X}=\left(x,y,z,\ell\right)$, which has its dynamics governed
by the Lagrangian density 
\begin{equation}
\mathcal{L}=\frac{1}{2}g^{MN}\partial_{M}\Phi\partial_{N}\Phi,\label{lagrangian}
\end{equation}
where $g^{MN}$ is the diagonal tensor metric given by the line element
(\ref{metric}), and capital Latin indices take values $0,1,2,3,4$.
The action which describes the scalar $\Phi$ field in this $5D$
spacetime, is given by
\begin{align}
S\left[\Phi\right] & =\int d^{5}x\:\sqrt{\left|g\right|}\:\mathcal{L}\:\delta\left(\ell-L\right)\nonumber \\
 & =\int d^{4}x\:d\ell\:\frac{\sqrt{A^{3}\left(t,\ell\right)B\left(t\right)}}{2\ell}\:\delta\left(\ell-L\right)\nonumber \\
 & \times\left\{ \left(\frac{\partial\Phi}{\partial t}\right)^{2}-\frac{\ell^{2}}{A\left(t,\ell\right)}\left[\left(\frac{\partial\Phi}{\partial x}\right)^{2}+\left(\frac{\partial\Phi}{\partial y}\right)^{2}+\left(\frac{\partial\Phi}{\partial z}\right)^{2}\right]-\frac{\ell^{2}}{B\left(t\right)}\left(\frac{\partial\Phi}{\partial\ell}\right)^{2}\right\} .\label{5}
\end{align}
Now, we must apply the cylinder condition on the field, i.e., $\partial\Phi/\partial\ell=0$,
and integrate over $\ell$-coordinate so that the Eq. (\ref{5}) assumes
the form
\begin{align}
S\left[\Phi\right] & =\int d^{4}x\:\frac{\sqrt{A^{3}\left(t,L\right)B\left(t\right)}}{2L}\nonumber \\
 & \times\left\{ \left(\frac{\partial\Phi_{\left(4\right)}}{\partial t}\right)^{2}-\frac{L^{2}}{A\left(t,L\right)}\left[\left(\frac{\partial\Phi_{\left(4\right)}}{\partial x}\right)^{2}+\left(\frac{\partial\Phi_{\left(4\right)}}{\partial y}\right)^{2}+\left(\frac{\partial\Phi_{\left(4\right)}}{\partial z}\right)^{2}\right]\right\} ,
\end{align}
where $\Phi_{\left(4\right)}=\left.\Phi\right|_{\ell=L}$. From this
point, we may decompose the scalar field into a complete basis, $u_{\mathbf{k}}=1/\sqrt{\mathcal{V}}e^{i\overrightarrow{k}\cdot\overrightarrow{R}}$,
normalized in a finite volume $\mathcal{V}$, with $\phi_{k}=\phi_{k}\left(t\right)$
written in terms of real and imaginary parts $\left(i=1,2\right)$,
so that
\begin{equation}
\Phi_{\left(4\right)}=\sum_{\overrightarrow{k}}\frac{1}{\sqrt{\mathcal{V}}}e^{i\overrightarrow{k}\cdot\overrightarrow{R}}\phi_{k}\left(t\right),\quad\phi_{k}\left(t\right)=\frac{1}{\sqrt{2}}\left(\phi_{k}^{1}+i\phi_{k}^{2}\right),\quad\overrightarrow{R}=\left(x,y,z\right),\overrightarrow{k}=\left(k_{x},k_{y},k_{z}\right).\label{dec}
\end{equation}
Since $4D$ coordinates of the spacetime are homogeneous and isotropic,
it is straightforward to obtain the Fourier transform of $\Phi_{\left(4\right)}$
(e.g., \citep{Pedrosa2021,Bertoni1998,Pedrosa2004,Alencar2012}),
such that the decomposition (\ref{dec}) yields the action for the
scalar field, 
\begin{equation}
S_{\phi}=\frac{1}{2}\sum_{k}\sum_{i=1,2}\int dt\frac{\sqrt{A^{3}\left(t,L\right)B\left(t\right)}}{L}\left[\dot{\phi}_{k}^{i\,2}-\omega_{k}^{2}\left(t,L\right)\phi_{k}^{i\,2}\right],\label{action2}
\end{equation}
in the background (\ref{metric}). Dot denote time derivative, and 
\begin{equation}
\omega_{k}^{2}\left(t,L\right)\equiv\frac{k^{2}L^{2}}{A\left(t,L\right)}=\omega_{0k}^{2}\left(\tau/t\right)^{2/\alpha},\text{ \ }\omega_{0k}=k\,L^{\frac{\alpha}{\alpha-1}}\tau^{-1/\alpha},\label{7}
\end{equation}
$\tau$ being the initial time.

Henceforward, to avoid notational clutter, we suppress the $i$-index
on the scalar field. No confusion will arise because $k$ and $i$
are independent indices \citep{Pedrosa2021,Bertoni1998}. It is now
straightforward to define a Hamiltonian for each $k$-mode of the
scalar field from the action (\ref{action2}), such that 
\begin{equation}
H_{k}=\frac{\Pi_{k}^{2}}{2m}+\frac{1}{2}m\omega_{k}^{2}\phi_{k}^{2},\label{hamiltonian}
\end{equation}
with 
\begin{equation}
\Pi_{k}=\frac{\partial L_{k}}{\partial\dot{\phi}_{k}}=\sqrt{\frac{A^{3}\left(t,L\right)B\left(t\right)}{L}}\dot{\phi}_{k},\text{ \ }m=m_{0}\left(t/\tau\right)^{\frac{3+\alpha}{\alpha}},\text{ \ }m_{0}=\frac{\alpha L^{\frac{5+\alpha}{2\left(1-\alpha\right)}}}{\left\vert \alpha-1\right\vert }\tau^{\frac{3+\alpha}{\alpha}},\label{9}
\end{equation}
where $\Pi_{k}$ is the conjugate momentum to the field $\phi_{k}$.
The classical equation of motion for $\phi_{k}$ can be obtained by
varying the action with respect to the field. Alternatively, turning
our attention to Eq. (\ref{hamiltonian}) we get the classical equation
of motion 
\begin{equation}
\ddot{\phi}_{k}+\frac{3+\alpha}{\alpha t}\dot{\phi}_{k}+\omega_{k}^{2}\phi_{k}=0,\label{EOM}
\end{equation}
whose solution is 
\begin{equation}
\phi_{k}\left(t\right)=C_{1}t^{-3/2\alpha}J_{\nu}\left(\frac{kt^{\frac{\alpha-1}{\alpha}}L^{\frac{\alpha}{\alpha-1}}\alpha}{\left\vert \alpha-1\right\vert }\right)+C_{2}t^{-3/2\alpha}Y_{\nu}\left(\frac{kt^{\frac{\alpha-1}{\alpha}}L^{\frac{\alpha}{\alpha-1}}\alpha}{\left\vert \alpha-1\right\vert }\right),\label{12}
\end{equation}
where $C_{1},C_{2}$ are real constants, $J_{\nu}$, and $Y_{\nu}$
are Bessel functions of first and second kind, respectively, and 
\begin{equation}
\nu=\frac{3}{2\left\vert \alpha-1\right\vert }.\label{13}
\end{equation}

Let us now identify the scalar field $\phi_{k}$ as an inflaton-like
field. To deal with this, it should be noted that, by assuming a coordinate
transformation such that $t=T^{\left(3-\alpha\right)/3}$, the Eq.
(\ref{EOM}) reduces to 
\begin{equation}
\ddot{\phi}_{k}+\frac{3}{\alpha T}\dot{\phi}_{k}+\widetilde{\omega}_{k}^{2}\phi_{k}=0,\label{14}
\end{equation}
where $\widetilde{\omega}_{k}=\frac{(\alpha-3)}{3}kT^{-\left(\alpha^{2}-\alpha+3\right)/3\alpha}L^{\alpha/\left(\alpha-1\right)}$,
and the overdot denotes partial derivative with respect to $T$. It
is important to highlight that from Eq. (\ref{14}), we can identify
the scalar field $\phi_{k}$ as a $4D$ massive inflaton field with
a quadratic potential 
\begin{equation}
V\left(\phi_{k}\right)=\frac{1}{2}\widetilde{\omega}_{k}^{2}\phi_{k}^{2}.\label{15}
\end{equation}

Therefore, Eq. (\ref{14}) characterizes the dynamics of the inflaton
field during inflation, radiation and matter eras. Furthermore, from
Eqs. (\ref{EOM}) and (\ref{14}), one can observe that the oscillation
modes of scalar waves in the time-dependent background can be analogously
described as a parametric oscillator system. In the following section,
we employ this analogy and apply it to the construction of SCSs for
the quantized scalar field.

\section{Quantum description of the scalar field\label{sec3}}

In this section, we aimed to analyze the propagating quantum scalar
field in a Kaluza-Klein cosmological background from a quantum mechanics
point-of-view. In its turn, the quantum description of the system
given by the Hamiltonian (\ref{hamiltonian}) can be provided through
canonical quantization, which in turn consists of taking $\Pi_{k}\rightarrow\hat{\Pi}_{k}$,
and $\phi_{k}\rightarrow\hat{\phi}_{k}$, i.e., we promote the scalar
field to an operator $\hat{\phi_{k}}$ that satisfies the canonical
commutation relations, 
\begin{equation}
\Pi_{k}\longrightarrow\hat{\Pi}_{k}=-i\hbar\frac{\partial}{\partial\phi_{k}},\text{ \ }\phi_{k}\longrightarrow\hat{\phi}_{k},\text{ \ }\left[\hat{\phi}_{k},\hat{\Pi}_{k^{\prime}}\right]=i\hbar\delta_{kk^{\prime}},\label{Q1}
\end{equation}
where $\hat{\Pi}_{k}$ is the conjugate momentum to $\hat{\phi}_{k}$,
and $\delta_{kk^{\prime}}$ is the Kronecker delta. On its turn, we
can represent these operators in terms of the canonical creation and
annihilation operators, $\hat{a}_{k}^{\dagger}$ and $\hat{a}_{k}$,
respectively, in the form 
\begin{equation}
\hat{\phi}_{k}=l_{k}\frac{\hat{a}_{k}+\hat{a}_{k}^{\dagger}}{\sqrt{2}},\text{ \ }\hat{\Pi}_{k}=\hbar\frac{\hat{a}_{k}-\hat{a}_{k}^{\dagger}}{i\sqrt{2}l_{k}},\text{ \ }\left[\hat{a}_{k},\hat{a}_{k}^{\dagger}\right]=1,\label{Q2}
\end{equation}
where $l_{k}$ is a parameter with length dimension. Thus, in terms
of the transformations (\ref{Q1}) and (\ref{Q2}), we can quantize
the Hamiltonian (\ref{hamiltonian}) into the Hamiltonian $\hat{H}_{k}$,
as follows 
\begin{equation}
\hat{H}_{k}=\frac{\hat{\Pi}_{k}^{2}}{2m}+\frac{1}{2}m\omega_{k}^{2}\hat{\phi}_{k}^{2}=\frac{\hbar\eta_{-}}{2}\left(\hat{a}_{k}^{2}+\hat{a}_{k}^{\dagger2}\right)+\hbar\eta_{+}\hat{a}_{k}^{\dagger}\hat{a}_{k}+\frac{\hbar\eta_{+}}{2},\label{Q3}
\end{equation}
where 
\begin{align}
 & \hat{\Pi}_{k}^{2}=\frac{\hbar^{2}}{2l_{k}^{2}}\left(\hat{a}_{k}^{\dagger}\hat{a}_{k}+\hat{a}_{k}\hat{a}_{k}^{\dagger}-\hat{a}_{k}^{2}-\hat{a}_{k}^{\dagger2}\right),\text{ \ }\hat{\phi}_{k}^{2}=\frac{l_{k}^{2}}{2}\left(\hat{a}_{k}^{\dagger}\hat{a}_{k}+\hat{a}_{k}\hat{a}_{k}^{\dagger}+\hat{a}_{k}^{2}+\hat{a}_{k}^{\dagger2}\right),\nonumber \\
 & \eta_{\pm}=\frac{l_{k}^{4}m^{2}\omega_{k}^{2}\pm\hbar^{2}}{2\hbar ml_{k}^{2}}=\frac{l_{k}^{2}m_{0}\omega_{0k}^{2}}{2\hbar}\left(t/\tau\right)^{\frac{1+\alpha}{\alpha}}\pm\frac{\hbar}{2l_{k}^{2}m_{0}}\left(t/\tau\right)^{-\frac{3+\alpha}{\alpha}}.\label{Q4}
\end{align}

So, the time evolution of the quantum states $\left\vert \Psi_{k}\right\rangle $
is given by the Schr\"{o}dinger equation 
\begin{align}
 & \hat{\Lambda}_{k}\left\vert \Psi_{k}\right\rangle =0,\label{Q5a}\\
 & \hat{\Lambda}_{k}=\hat{H}_{k}-i\hbar\partial_{t}.\label{Q5b}
\end{align}

We can obtain the solution to (\ref{Q5a}) by applying the integral
of motion method \citep{Lewis1969,Dodonov1975,Dodonov2003}, which
consists of defining a time-dependent operator, for instance, $\hat{A}_{k}\left(t\right)$,
which commutes with $\hat{\Lambda}_{k}$, such that 
\begin{equation}
\frac{d\hat{A}_{k}}{dt}=\frac{i}{\hbar}\left[\hat{\Lambda}_{k},\hat{A}_{k}\right]=0.\label{Q6}
\end{equation}
The eigenstates of $\hat{A}_{k}\left(t\right)$ must satisfy equation
(\ref{Q5a}). Following the proposal presented in \citep{Pereira2021,Pereira2022,Pereira2023},
we will define $\hat{A}_{k}\left(t\right)$ in the form of a linear
combination of the operators $\hat{a}_{k}$ and $\hat{a}_{k}^{\dagger}$,
yielding to 
\begin{align}
 & \hat{A}_{k}=f_{k}\hat{a}_{k}+g_{k}\hat{a}_{k}^{\dagger}+\varphi_{k},\nonumber \\
 & \left[\hat{A}_{k},\hat{A}_{k}^{\dagger}\right]=\mu_{k},\text{ \ }\mu_{k}\equiv\left\vert f_{k}\right\vert ^{2}-\left\vert g_{k}\right\vert ^{2},\label{Q7}
\end{align}
where $f_{k}=f_{k}\left(t\right)$, $g_{k}=g_{k}\left(t\right)$,
and $\varphi_{k}=\varphi_{k}\left(t\right)$ are time-dependent complex
functions.

Replacing the representations (\ref{Q3}), (\ref{Q5b}), and (\ref{Q7})
into (\ref{Q6}), we get 
\begin{align}
 & \dot{g}_{k}=i\eta_{-}f_{k}-i\eta_{+}g_{k},\nonumber \\
 & \dot{f}_{k}=i\eta_{+}f_{k}-i\eta_{-}g_{k},\nonumber \\
 & \dot{\varphi}_{k}=0.\label{Q8}
\end{align}
Given the initial conditions $f_{0k}=f_{k}\left(\tau\right)$, $g_{0k}=g_{k}\left(\tau\right)$,
and $\varphi_{0k}=\varphi_{k}\left(\tau\right)$, we will have the
following solution for the system (\ref{Q8}), 
\begin{align}
f_{k}= & \frac{\pi\beta_{k}\tau^{\delta}}{4}\left[\frac{\tau^{\delta\left(\varepsilon-1\right)}}{t^{\delta\left(\varepsilon-1\right)}}W_{k,\varepsilon,\varepsilon-1}^{\left(\tau,t\right)}F_{0k}+\frac{t^{\delta\varepsilon}}{\tau^{\delta\varepsilon}}W_{k,\varepsilon,\varepsilon-1}^{\left(t,\tau\right)}G_{0k}\right]\nonumber \\
+ & \frac{i\pi\hbar\left(\tau t\right)^{\delta\varepsilon+\frac{3+\alpha}{\alpha}}}{4l_{k}^{2}\delta m}\left[W_{k,\varepsilon,\varepsilon}^{\left(\tau,t\right)}F_{0k}-\frac{l_{k}^{4}\delta^{2}\beta_{k}^{2}m^{2}}{\hbar^{2}\left(\tau t\right)^{2\delta\varepsilon+\frac{7+\alpha}{\alpha}}}W_{k,\varepsilon-1,\varepsilon-1}^{\left(t,\tau\right)}G_{0k}\right],\nonumber \\
g_{k}= & \frac{\pi\beta_{k}\tau^{\delta}}{4}\left[\frac{\tau^{\delta\left(\varepsilon-1\right)}}{t^{\delta\left(\varepsilon-1\right)}}W_{k,\varepsilon,\varepsilon-1}^{\left(\tau,t\right)}F_{0k}-\frac{t^{\delta\varepsilon}}{\tau^{\delta\varepsilon}}W_{k,\varepsilon,\varepsilon-1}^{\left(t,\tau\right)}G_{0k}\right]\nonumber \\
- & \frac{i\pi}{4}\frac{\hbar\left(\tau t\right)^{\delta\varepsilon+\frac{3+\alpha}{\alpha}}}{l_{k}^{2}\delta m}\left[W_{k,\varepsilon,\varepsilon}^{\left(\tau,t\right)}F_{0k}+\frac{l_{k}^{4}\delta^{2}\beta_{k}^{2}m^{2}}{\hbar^{2}\left(\tau t\right)^{2\delta\varepsilon+\frac{7+\alpha}{\alpha}}}W_{k,\varepsilon-1,\varepsilon-1}^{\left(t,\tau\right)}G_{0k}\right],\label{Q9}
\end{align}
where $F_{0k}$, $G_{0k}$ are constants determined by the initial
conditions, and $W_{k,a,b}^{\left(\tau,t\right)}=J_{a}\left(\beta_{k}\tau^{\delta}\right)Y_{b}\left(\beta_{k}t^{\delta}\right)-J_{b}\left(\beta_{k}t^{\delta}\right)Y_{a}\left(\beta_{k}\tau^{\delta}\right)$.
For more details, please see Appendix \ref{Appendix-A}.

From here, it is convenient to define a new operator, in the form
\begin{equation}
\hat{B}_{k}\equiv\frac{1}{f_{k}}\hat{A}_{k}=\hat{a}_{k}+\zeta_{k}\hat{a}_{k}^{\dagger}+\xi_{k},\text{ \ }\left[\hat{B}_{k},\hat{B}_{k}^{\dagger}\right]=1-\left\vert \zeta_{k}\right\vert ^{2},\label{Q10}
\end{equation}
where the time-dependent quantities $\xi_{k}$ and $\zeta_{k}$, are
given by 
\begin{equation}
\xi_{k}\equiv\frac{\varphi_{k}}{f_{k}},\text{ \ }\zeta_{k}\equiv\frac{g_{k}}{f_{k}},\label{Q11}
\end{equation}
which are identified as displacement and squeeze parameters, respectively,
in the non-unitary treatment of SCSs. In what follows, we can write
$\hat{a}_{k}$ and $\hat{a}_{k}^{\dagger}$ in terms of the new operators
$\hat{B}_{k}$ and $\hat{B}_{k}^{\dagger}$, whose explicitly forms
are 
\begin{equation}
\hat{a}_{k}=\frac{\hat{B}_{k}-\zeta_{k}\hat{B}_{k}^{\dagger}+\zeta_{k}\xi_{k}^{\ast}-\xi_{k}}{1-\left\vert \zeta_{k}\right\vert ^{2}},\text{ \ }\hat{a}_{k}^{\dagger}=\frac{\hat{B}_{k}^{\dagger}-\zeta_{k}^{\ast}\hat{B}_{k}+\zeta_{k}^{\ast}\xi_{k}-\xi_{k}^{\ast}}{1-\left\vert \zeta_{k}\right\vert ^{2}}.\label{Q12}
\end{equation}
As we will see, from relations (\ref{Q12}), we can obtain the average
values and uncertainty relation to the built SCSs.

\section{Construction of SCSs in a non-unitary approach\label{sec4}}

The employ of quantum optics techniques, as the SSs formalism, has
proven highly advantageous for addressing numerous issues in cosmology
\citep{Grishchuk1989,Grishchuk1990,Pedrosa2021,Bertoni1998,Pedrosa2004,Suresh2004}.
In its turn, the integrals of motion in a non-unitary approach has
shown itself as a suitable method for the study of time-dependent
quantum systems \citep{Andrews1999,Pereira2021,Pereira2022,Pereira2023}.
Therefore, by applying the non-unitary treatment, we intend to construct
the SCSs for this system. So, let us first define a non-unitary operator
composed by the displacement ($\xi_{k}$) and squeeze ($\zeta_{k}$)
parameters, as we see below 
\begin{equation}
\hat{S}_{k}=\exp\left(\xi_{k}\hat{a}_{k}^{\dagger}+\frac{1}{2}\zeta_{k}\hat{a}_{k}^{\dagger2}\right).\label{W1}
\end{equation}
By applying the Baker--Campbell--Hausdorff theorem, i.e., $e^{A}Be^{-A}=B+\left[A,B\right]+2^{-1}\left[A,\left[A,B\right]\right]+...$,
it is possible to transform $\hat{B}_{k}$ into the operator $\hat{a}_{k}$,
by considering the relation 
\begin{equation}
\hat{a}_{k}=\hat{S}_{k}\hat{B}_{k}\hat{S}_{k}^{-1}.\label{W2}
\end{equation}
So, by applying Eq. (\ref{W2}) on the vacuum state $\left\vert 0_{k}\right\rangle $,
which is annihilated by $\hat{a}_{k}$, i.e. $\hat{a}_{k}\left\vert 0_{k}\right\rangle =0$,
we obtain 
\begin{equation}
\hat{B}_{k}\left\vert \xi_{k},\zeta_{k}\right\rangle =0,\text{ \ }\left\vert \xi_{k},\zeta_{k}\right\rangle =\Phi_{k}\exp\left(-\xi_{k}\hat{a}_{k}^{\dagger}-\frac{1}{2}\zeta_{k}\hat{a}_{k}^{\dagger2}\right)\left\vert 0_{k}\right\rangle ,\label{W3}
\end{equation}
where $\Phi_{k}=\Phi_{k}\left(t\right)$ is a function to be determined
such that the states $\left\vert \xi_{k},\zeta_{k}\right\rangle $
simultaneously satisfy the normalization condition and Eq. (\ref{Q5a}).

From the normalization condition applied to the states $\left\vert \xi_{k},\zeta_{k}\right\rangle $,
that is, $\left\langle \zeta_{k},\xi_{k}|\xi_{k},\zeta_{k}\right\rangle =1$,
we obtain, 
\begin{equation}
\left\vert \Phi_{k}\right\vert =\left(1-\left\vert \zeta_{k}\right\vert ^{2}\right)^{1/4}\exp\left[-\frac{1}{2}\frac{\left\vert \xi_{k}\right\vert ^{2}-\operatorname{Re}\left(\zeta_{k}\xi_{k}^{\ast2}\right)}{1-\left\vert \zeta_{k}\right\vert ^{2}}\right].\label{W5}
\end{equation}
In order to find the previous result, we used the relations below:
\begin{align}
 & \exp\left(2yz-z^{2}\right)=%TCIMACRO{\dsum\limits _{n=0}^{\infty}}%
%BeginExpansion
{\displaystyle \sum\limits _{n=0}^{\infty}}%EndExpansion
\frac{H\left(y\right)}{n!}z^{n},\text{ \ }z=-\sqrt{\frac{\zeta_{k}}{2}}\hat{a}_{k}^{\dagger},\text{ \ }y=\frac{\xi_{k}}{\sqrt{2\zeta_{k}}},\nonumber \\
 & %TCIMACRO{\dsum\limits _{n=0}^{\infty}}%
%BeginExpansion
{\displaystyle \sum\limits _{n=0}^{\infty}}%EndExpansion
H_{n}\left(x\right)H_{n}\left(y\right)\frac{z^{n}}{2^{n}n!}=\frac{1}{\sqrt{1-z^{2}}}\exp\left[\frac{2xyz-\left(x^{2}+y^{2}\right)z^{2}}{1-z^{2}}\right],\nonumber \\
 & x=\frac{\xi_{k}^{\ast}}{\sqrt{2\zeta_{k}^{\ast}}},\text{ \ }y=\frac{\xi_{k}}{\sqrt{2\zeta_{k}}},\text{ \ }z=\left\vert \zeta_{k}\right\vert ,
\end{align}
where $H_{n}\left(x\right)$ are the Hermite polynomials. Thus, given
that $\Phi_{k}=\left\vert \Phi_{k}\right\vert e^{i\theta_{k}}$, the
normalized states can be written as 
\begin{equation}
\left\vert \xi_{k},\zeta_{k}\right\rangle =\left(1-\left\vert \zeta_{k}\right\vert ^{2}\right)^{1/4}\exp\left(-\frac{1}{2}\frac{\left\vert \xi_{k}\right\vert ^{2}-\zeta_{k}^{\ast}\xi_{k}^{2}}{1-\left\vert \zeta_{k}\right\vert ^{2}}+i\theta_{k}\right)\exp\left(-\xi_{k}\hat{a}_{k}^{\dagger}-\frac{1}{2}\zeta_{k}\hat{a}_{k}^{\dagger2}\right)\left\vert 0_{k}\right\rangle .\label{W6}
\end{equation}
To determine $\theta_{k}$ we assumed that the states $\left\vert \xi_{k},\zeta_{k}\right\rangle $
satisfy equation (\ref{Q5a}). For this purpose, we must solve the
equation below, 
\begin{equation}
\left\langle \zeta_{k},\xi_{k}\left\vert \hat{H}_{k}-i\hbar\partial_{t}\right\vert \xi_{k},\zeta_{k}\right\rangle =0.\label{W7}
\end{equation}
Since 
\begin{equation}
\left\langle \zeta_{k},\xi_{k}\left\vert \hat{H}_{k}\right\vert \xi_{k},\zeta_{k}\right\rangle =i\hbar\left\langle \zeta_{k},\xi_{k}\left\vert \partial_{t}\right\vert \xi_{k},\zeta_{k}\right\rangle +\hbar\dot{\theta}_{k}-\frac{\hbar\eta_{-}}{2}\operatorname{Re}\left(\zeta_{k}\right)+\frac{\hbar\eta_{+}}{2},\label{W8}
\end{equation}
where 
\begin{align}
\left\langle \zeta_{k},\xi_{k}\left\vert \hat{H}_{k}\right\vert \xi_{k},\zeta_{k}\right\rangle = & \frac{\hbar\eta_{-}}{2}\frac{\left(\zeta_{k}\xi_{k}^{\ast}-\xi_{k}\right)^{2}+\left(\zeta_{k}^{\ast}\xi_{k}-\xi_{k}^{\ast}\right)^{2}-\left(\zeta_{k}+\zeta_{k}^{\ast}\right)\left(1-\left\vert \zeta_{k}\right\vert ^{2}\right)}{\left(1-\left\vert \zeta_{k}\right\vert ^{2}\right)^{2}}\nonumber \\
+ & \hbar\eta_{+}\frac{\left\vert \zeta_{k}\xi_{k}^{\ast}-\xi_{k}\right\vert ^{2}+\left(1-\left\vert \zeta_{k}\right\vert ^{2}\right)\left\vert \zeta_{k}\right\vert ^{2}}{\left(1-\left\vert \zeta_{k}\right\vert ^{2}\right)^{2}}+\frac{\hbar\eta_{+}}{2},\label{W9}
\end{align}
and, by imposing that the states $\left\vert \xi_{k},\zeta_{k}\right\rangle $
satisfy the Schr\"{o}dinger equation, we must have, 
\begin{equation}
\dot{\theta}_{k}=\frac{1}{2}\operatorname{Re}\left(\eta_{-}\zeta_{k}-\eta_{+}\right)\Longrightarrow\theta_{k}=\frac{1}{2}\int_{\tau}^{t}\operatorname{Re}\left(\eta_{-}\zeta_{k}-\eta_{+}\right)dt^{\prime}.\label{W10}
\end{equation}
Finally, the normalized states satisfying equations (\ref{Q5a}) and
(\ref{W3}) take the form 
\begin{equation}
\left\vert \xi_{k},\zeta_{k}\right\rangle =\left(1-\left\vert \zeta_{k}\right\vert ^{2}\right)^{1/4}\exp\left(\frac{1}{2}\frac{\zeta_{k}^{\ast}\xi_{k}^{2}-\left\vert \xi_{k}\right\vert ^{2}}{1-\left\vert \zeta_{k}\right\vert ^{2}}+i\theta_{k}\right)%TCIMACRO{\dsum\limits _{n_{k}=0}^{\infty}}%
%BeginExpansion
{\displaystyle \sum\limits _{n_{k}=0}^{\infty}}%EndExpansion
\frac{\left(-1\right)^{n_{k}}}{\sqrt{n_{k}!}}\left(\frac{\zeta_{k}}{2}\right)^{\frac{n_{k}}{2}}H_{n_{k}}\left(\frac{\xi_{k}}{\sqrt{2\zeta_{k}}}\right)\left\vert n_{k}\right\rangle .\label{W11}
\end{equation}
Thus, we have built the SCSs for the quantum field immersed in the
cosmological Kaluza-Klein background, whose dynamics is described
by Hamiltonian (\ref{Q3}). The physical properties of these states
will be investigated in the following subsections.

Besides finding the normalized states, we can also determine the transition
probability $P_{n_{k}}=\left\vert \left\langle n_{k}|\zeta_{k},\xi_{k}\right\rangle \right\vert ^{2}$.
In this case, we assume that the system was initially prepared in
a time-independent Fock state and will suffer a transition to SCSs.
Thus, we now can use the results (\ref{W11}) in order to obtain the
transition probability as follows: 
\begin{equation}
P_{n_{k}}=\sqrt{1-\left\vert \zeta_{k}\right\vert ^{2}}\exp\left[\frac{\operatorname{Re}\left(\zeta_{k}^{\ast}\xi_{k}^{2}\right)-\left\vert \xi_{k}\right\vert ^{2}}{1-\left\vert \zeta_{k}\right\vert ^{2}}\right]\frac{\left\vert \zeta_{k}\right\vert ^{n_{k}}}{2^{n_{k}}n_{k}!}\left\vert H_{n_{k}}\left(\frac{\xi_{k}}{\sqrt{2\zeta_{k}}}\right)\right\vert ^{2}.\label{R7-1}
\end{equation}
Our next subsection will be dedicated to discuss the quadrature squeezing
properties of our system.

\subsection{Quadrature squeezing properties: Mean values and minimization of
uncertainty relation}

In this subsection, in order to investigate the non-classicality of
constructed SCSs, we will analyze the quadrature squeezing properties
from uncertainty relations. Therefore, let us start by considering
$\hat{\phi}_{k}$ and $\hat{\Pi}_{k}$ given in terms of the integrals
of motion: 
\begin{align}
 & \hat{\phi}_{k}=\frac{l_{k}}{\sqrt{2}}\frac{\left(1-\zeta_{k}^{\ast}\right)\hat{B}_{k}+\left(1-\zeta_{k}\right)\hat{B}_{k}^{\dagger}-2\operatorname{Re}\left[\left(1-\zeta_{k}^{\ast}\right)\xi_{k}\right]}{1-\left\vert \zeta_{k}\right\vert ^{2}},\nonumber \\
 & \hat{\Pi}_{k}=\frac{\hbar}{\sqrt{2}il_{k}}\frac{\left(1+\zeta_{k}^{\ast}\right)\hat{B}_{k}-\left(1+\zeta_{k}\right)\hat{B}_{k}^{\dagger}-2i\operatorname{Im}\left[\left(1+\zeta_{k}^{\ast}\right)\xi_{k}\right]}{1-\left\vert \zeta_{k}\right\vert ^{2}}.\label{M1}
\end{align}
where $l_{k}$ is a length-dimensional parameter which is related
to the initial standard deviation \citep{Pereira2021}. Using the
relations (\ref{W3}) and (\ref{M1}), we can easily calculate the
mean values of the operators $\hat{\phi}_{k}$ and $\hat{\Pi}_{k}$,
whose explicitly forms are 
\begin{align}
\overline{\phi_{k}} & =\overline{\phi_{k}}\left(t\right)=\left\langle \xi_{k},\zeta_{k}\left\vert \hat{\phi}_{k}\right\vert \zeta_{k},\xi_{k}\right\rangle =-\frac{\sqrt{2}l_{k}\operatorname{Re}\left[\left(1-\zeta_{k}^{\ast}\right)\xi_{k}\right]}{1-\left\vert \zeta_{k}\right\vert ^{2}},\nonumber \\
\overline{\Pi_{k}} & =\overline{\Pi_{k}}\left(t\right)=\left\langle \xi_{k},\zeta_{k}\left\vert \hat{\Pi}_{k}\right\vert \zeta_{k},\xi_{k}\right\rangle =-\frac{\hbar}{l_{k}}\frac{\sqrt{2}\operatorname{Im}\left[\left(1+\zeta_{k}^{\ast}\right)\xi_{k}\right]}{1-\left\vert \zeta_{k}\right\vert ^{2}}.\label{M2}
\end{align}
From here, we can identify that there is a correspondence between
the squeeze $\zeta_{k}$ and displacement $\xi_{k}$ parameters with
the mean values of $\overline{\phi_{k}}$ and $\overline{\Pi_{k}}$,
such that 
\begin{equation}
\xi_{k}=-\frac{1+\zeta_{k}}{\sqrt{2}l_{k}}\overline{\phi_{k}}-\frac{il_{k}}{\hbar}\frac{1-\zeta_{k}}{\sqrt{2}}\overline{\Pi_{k}}.\label{M3}
\end{equation}

Besides by taking the square of $\hat{\phi}_{k}$ and $\hat{\Pi}_{k}$,
we have 
\begin{align}
\hat{\phi}_{k}^{2} & =\frac{l_{k}^{2}}{2}\frac{\left(1-\zeta_{k}^{\ast}\right)^{2}\hat{B}_{k}^{2}+2\left\vert 1-\zeta_{k}\right\vert ^{2}\hat{B}_{k}^{\dagger}\hat{B}_{k}+\left(1-\zeta_{k}\right)^{2}\hat{B}_{k}^{\dagger2}}{\left(1-\left\vert \zeta_{k}\right\vert ^{2}\right)^{2}}\nonumber \\
 & -\frac{2l_{k}^{2}\operatorname{Re}\left[\left(1-\zeta_{k}^{\ast}\right)\xi_{k}\right]\left[\left(1-\zeta_{k}^{\ast}\right)\hat{B}_{k}+\left(1-\zeta_{k}\right)\hat{B}_{k}^{\dagger}\right]}{\left(1-\left\vert \zeta_{k}\right\vert ^{2}\right)^{2}}+\frac{l_{k}^{2}}{2}\frac{\left\vert 1-\zeta_{k}\right\vert ^{2}}{1-\left\vert \zeta_{k}\right\vert ^{2}}+\frac{2l_{k}^{2}\operatorname{Re}^{2}\left[\left(1-\zeta_{k}^{\ast}\right)\xi_{k}\right]}{\left(1-\left\vert \zeta_{k}\right\vert ^{2}\right)^{2}},\nonumber \\
\hat{\Pi}_{k}^{2} & =\frac{2i\hbar^{2}\operatorname{Im}\left[\left(1+\zeta_{k}^{\ast}\right)\xi_{k}\right]\left[\left(1+\zeta_{k}^{\ast}\right)\hat{B}_{k}-\left(1+\zeta_{k}\right)\hat{B}_{k}^{\dagger}\right]}{l_{k}^{2}\left(1-\left\vert \zeta_{k}\right\vert ^{2}\right)^{2}}\nonumber \\
 & -\frac{\hbar^{2}}{2l_{k}^{2}}\frac{\left(1+\zeta_{k}^{\ast}\right)^{2}\hat{B}_{k}^{2}-2\left\vert 1+\zeta_{k}\right\vert ^{2}\hat{B}_{k}^{\dagger}\hat{B}_{k}+\left(1+\zeta_{k}\right)^{2}\hat{B}_{k}^{\dagger2}}{\left(1-\left\vert \zeta_{k}\right\vert ^{2}\right)^{2}}+\frac{\hbar^{2}}{2l_{k}^{2}}\frac{\left\vert 1+\zeta_{k}\right\vert ^{2}}{1-\left\vert \zeta_{k}\right\vert ^{2}}+\frac{2\hbar^{2}\operatorname{Im}^{2}\left[\left(1+\zeta_{k}^{\ast}\right)\xi_{k}\right]}{l_{k}^{2}\left(1-\left\vert \zeta_{k}\right\vert ^{2}\right)^{2}}.\label{M4}
\end{align}
So, the previous results allow us to compute the mean values of $\overline{\phi_{k}^{2}}$
and $\overline{\Pi_{k}^{2}}$, as follows 
\begin{align}
\overline{\phi_{k}^{2}} & =\overline{\phi_{k}^{2}}\left(t\right)=\left\langle \xi_{k},\zeta_{k}\left\vert \hat{\phi}_{k}^{2}\right\vert \zeta_{k},\xi_{k}\right\rangle =\frac{l_{k}^{2}}{2}\frac{\left\vert 1-\zeta_{k}\right\vert ^{2}}{1-\left\vert \zeta_{k}\right\vert ^{2}}+\frac{2l_{k}^{2}\operatorname{Re}^{2}\left[\left(1-\zeta_{k}^{\ast}\right)\xi_{k}\right]}{\left(1-\left\vert \zeta_{k}\right\vert ^{2}\right)^{2}},\nonumber \\
\overline{\Pi_{k}^{2}} & =\overline{\Pi_{k}^{2}}\left(t\right)=\left\langle \xi_{k},\zeta_{k}\left\vert \hat{\Pi}_{k}^{2}\right\vert \zeta_{k},\xi_{k}\right\rangle =\frac{\hbar^{2}}{2l_{k}^{2}}\frac{\left\vert 1+\zeta_{k}\right\vert ^{2}}{1-\left\vert \zeta_{k}\right\vert ^{2}}+\frac{2\hbar^{2}\operatorname{Im}^{2}\left[\left(1+\zeta_{k}^{\ast}\right)\xi_{k}\right]}{l_{k}^{2}\left(1-\left\vert \zeta_{k}\right\vert ^{2}\right)^{2}}.\label{M5}
\end{align}
Moreover, we are also able to calculate the standard deviation, i.e.,
\begin{align}
 & \sigma_{\phi_{k}}=\sigma_{\phi_{k}}\left(t\right)=\sqrt{\overline{\phi_{k}^{2}}-\overline{\phi_{k}}^{2}}=\frac{l_{k}}{\sqrt{2}}\frac{\left\vert 1-\zeta_{k}\right\vert }{\sqrt{1-\left\vert \zeta_{k}\right\vert ^{2}}},\nonumber \\
 & \sigma_{\Pi_{k}}=\sigma_{\Pi_{k}}\left(t\right)=\sqrt{\overline{\Pi_{k}^{2}}-\overline{\Pi_{k}}^{2}}=\frac{\hbar}{\sqrt{2}l_{k}}\frac{\left\vert 1+\zeta_{k}\right\vert }{\sqrt{1-\left\vert \zeta_{k}\right\vert ^{2}}}.\label{M6}
\end{align}
From Eqs. (\ref{M6}), one can see that standard deviations associated
to the field and momentum present the squeezing property.

Finally, one can find the Heisenberg uncertainty relation to the SCSs.
Therefore, in the hold of previous results, we can explicitly calculate
the product $\sigma_{\phi_{k}}\sigma_{\Pi_{k}}$, which results in
\begin{equation}
\sigma_{\phi_{k}}\sigma_{\Pi_{k}}=\frac{\hbar}{2}\frac{\left\vert 1-\zeta_{k}\right\vert \left\vert 1+\zeta_{k}\right\vert }{1-\left\vert \zeta_{k}\right\vert ^{2}}=\frac{\hbar}{2}\sqrt{1+\frac{4}{\hbar^{2}}\sigma_{\phi_{k}\Pi_{k}}^{2}},\label{M7}
\end{equation}
where $\sigma_{\phi_{k}\Pi_{k}}$ is the covariance given by 
\begin{equation}
\sigma_{\phi_{k}\Pi_{k}}=\frac{\left\langle \zeta_{k},\xi_{k}\left\vert \hat{\Pi}_{k}\hat{\phi}_{k}+\hat{\phi}_{k}\hat{\Pi}_{k}\right\vert \xi_{k},\zeta_{k}\right\rangle }{2}-\overline{\phi_{k}}\left(t\right)\overline{\Pi_{k}}\left(t\right)=-\hbar\frac{\operatorname{Im}\left(\zeta_{k}\right)}{1-\left\vert \zeta_{k}\right\vert ^{2}}.\label{M8}
\end{equation}In turn, the Robertson-Schr\"odinger uncertainty relation can be calculated as follows,
\begin{equation}
\sigma_{\phi_{k}}^2\sigma_{\Pi_{k}}^2-\sigma_{\phi_{k}\Pi_{k}}^2=\frac{\hbar}{4}.
\end{equation}This result indicates that the SCS for the field corresponds to a class of correlated CS \citep{Dodonov1980}.

Now, let us consider the minimization of the uncertainty relation
at $t=\tau$. For this, admit that 
\begin{equation}
\zeta_{0k}=\zeta_{k}\left(\tau\right)=\tanh\left(r\right)\Rightarrow f_{0k}=\cosh\left(r\right),\text{ \ }g_{0k}=\sinh\left(r\right),\text{ \ }r\in\left(-\infty,\infty\right),\label{M8a}
\end{equation}
where the $r$-parameter is related to the magnitude of squeeze parameter
in the initial time. Besides, the conditions (\ref{M8a}) ensures
that $\mu_{0k}=\left\vert f_{0k}\right\vert ^{2}-\left\vert g_{0k}\right\vert ^{2}=1$.
Our next step consists in determining the parameter $l_{k}$ in terms
of $r$ and $\sigma_{\phi_{0k}}=\sigma_{\phi_{k}}\left(\tau\right)$.
So, by taking the relation (\ref{M6}) with $t=\tau$, we directly
observe that 
\begin{equation}
l_{k}=\sqrt{2}e^{r}\sigma_{\phi_{0k}}.\label{M9}
\end{equation}
Then, from (\ref{M8}), and (\ref{M8a}), we have $\sigma_{\phi_{k}\Pi_{k}}\left(\tau\right)=0$,
which implies that the relation (\ref{M7}) is minimized at $t=\tau$,
i.e. 
\begin{equation}
\sigma_{\phi_{0k}}\sigma_{\Pi_{0k}}=\frac{\hbar}{2},\text{ \ }\sigma_{\Pi_{0k}}=\sigma_{\Pi_{k}}\left(\tau\right).\label{M10}
\end{equation}
The condition found in Eq. (\ref{M9}) is going to by considered in
our future approaches.

Here, we intend to rewrite the uncertainty relations (\ref{M6}),
aiming to show the quadrature squeezing properties. For this, use
the following dimensionless transformation 
\begin{equation}
\hat{Q}_{k}=\frac{1}{l_{k}}\hat{\phi}_{k}=\frac{\hat{\phi}_{k}}{\sqrt{2}e^{r}\sigma_{\phi_{0k}}},\text{ \ }\hat{P}_{k}=\frac{l_{k}}{\hbar}\hat{\Pi}_{k}=\frac{\sqrt{2}e^{r}\sigma_{\phi_{0k}}}{\hbar}\hat{\Pi}_{k},\text{ \ }\left[\hat{Q}_{k},\hat{P}\right]=i.\label{M11}
\end{equation}
From here, we can rewrite (\ref{M6}) in the form 
\begin{align}
 & \sigma_{Q_{k}}=\frac{1}{l_{k}}\sigma_{\phi_{k}}=\frac{1}{\sqrt{2}}\frac{\left\vert 1-\zeta_{k}\right\vert }{\sqrt{1-\left\vert \zeta_{k}\right\vert ^{2}}},\nonumber \\
 & \sigma_{P_{k}}=\frac{l_{k}}{\hbar}\sigma_{\Pi_{k}}=\frac{1}{\sqrt{2}}\frac{\left\vert 1+\zeta_{k}\right\vert }{\sqrt{1-\left\vert \zeta_{k}\right\vert ^{2}}}.\label{M12}
\end{align}
So, the Heisenberg uncertainty relation becomes 
\begin{equation}
\sigma_{Q_{k}}\sigma_{P_{k}}=\frac{1}{2}\sqrt{1+\frac{4}{\hbar^{2}}\sigma_{\phi_{k}\Pi_{k}}^{2}},\text{ \ }\sigma_{\phi_{k}\Pi_{k}}^{2}=\frac{\hbar^{2}\operatorname{Im}^{2}\left(\zeta_{k}\right)}{\left(1-\left\vert \zeta_{k}\right\vert ^{2}\right)^{2}}.\label{M13}
\end{equation}

\begin{figure}[ptb]
\subfloat{\includegraphics[scale=0.35]{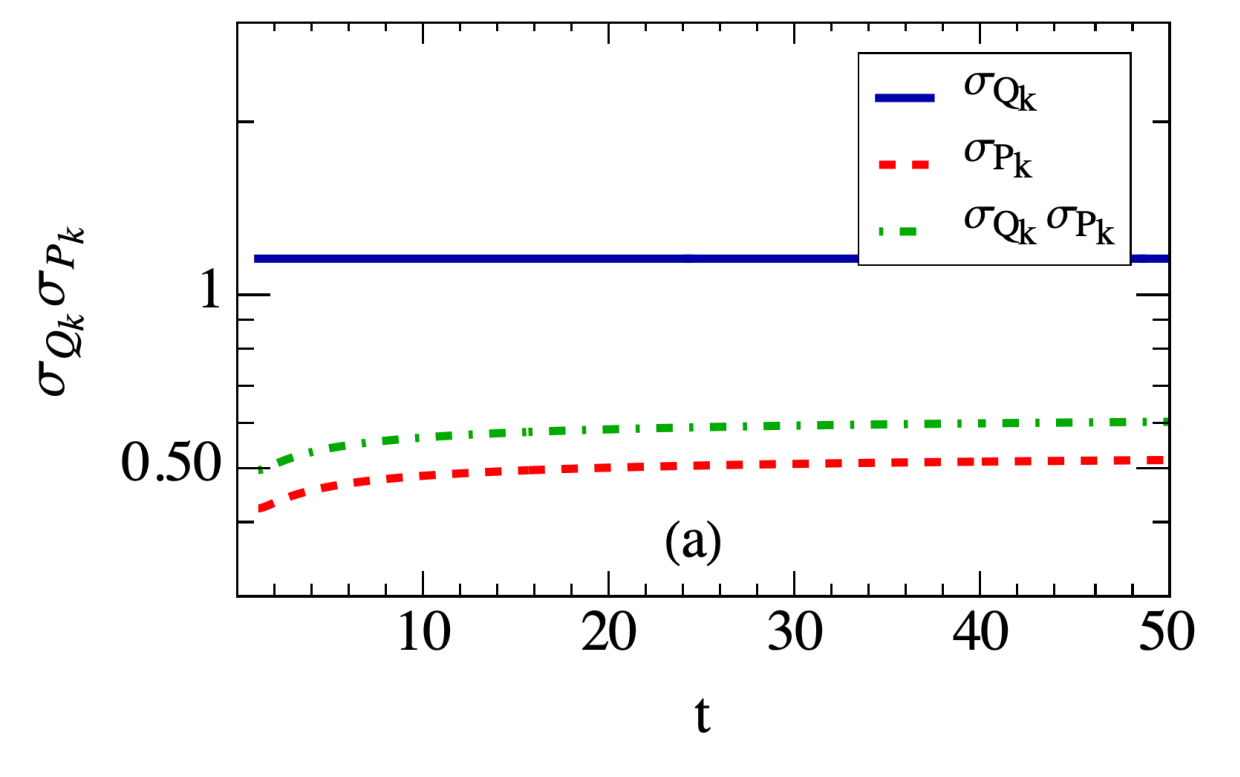}}\hfill{}\subfloat{\includegraphics[scale=0.35]{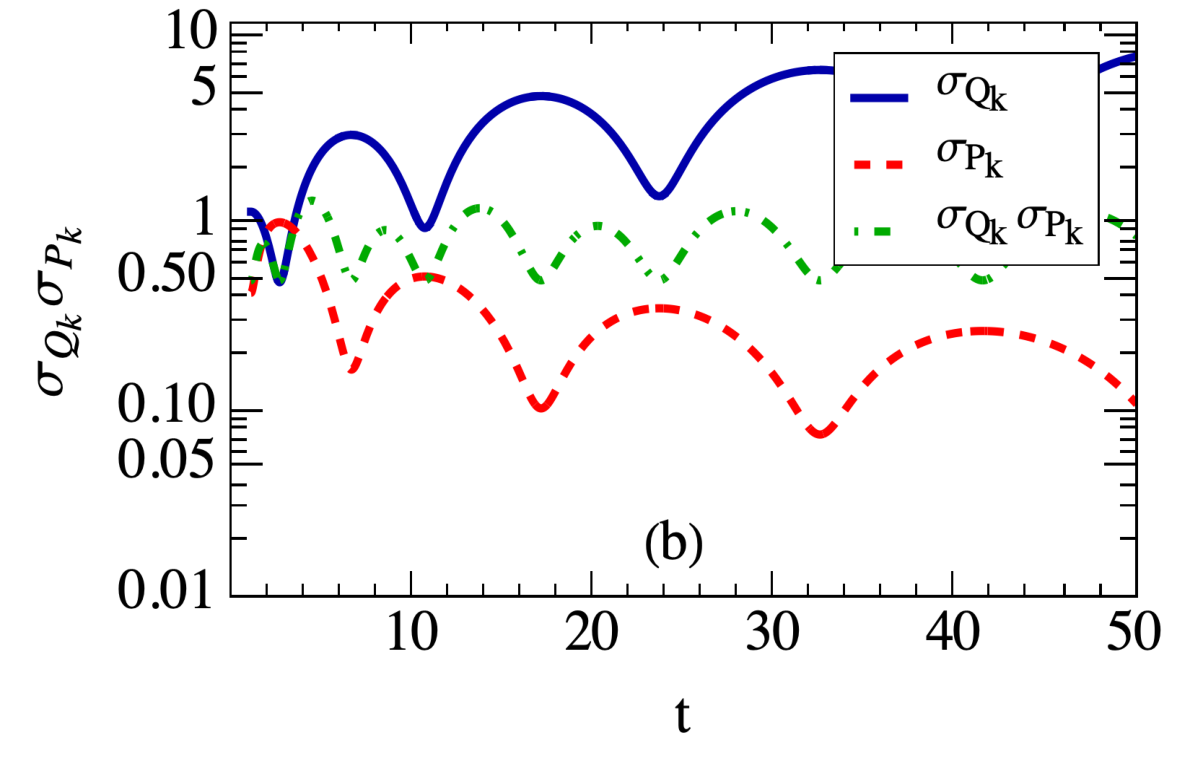}}\vfill{}
 \subfloat{\includegraphics[scale=0.35]{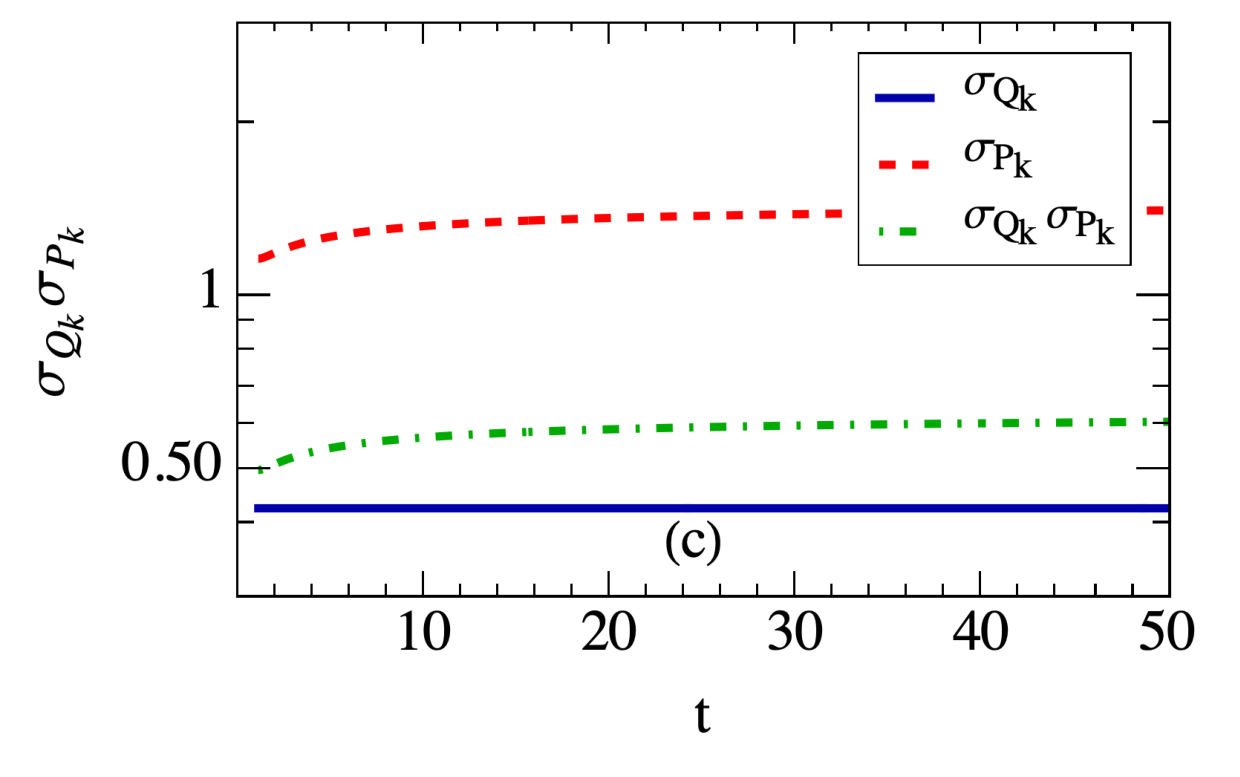}}\hfill{}\subfloat{\includegraphics[scale=0.35]{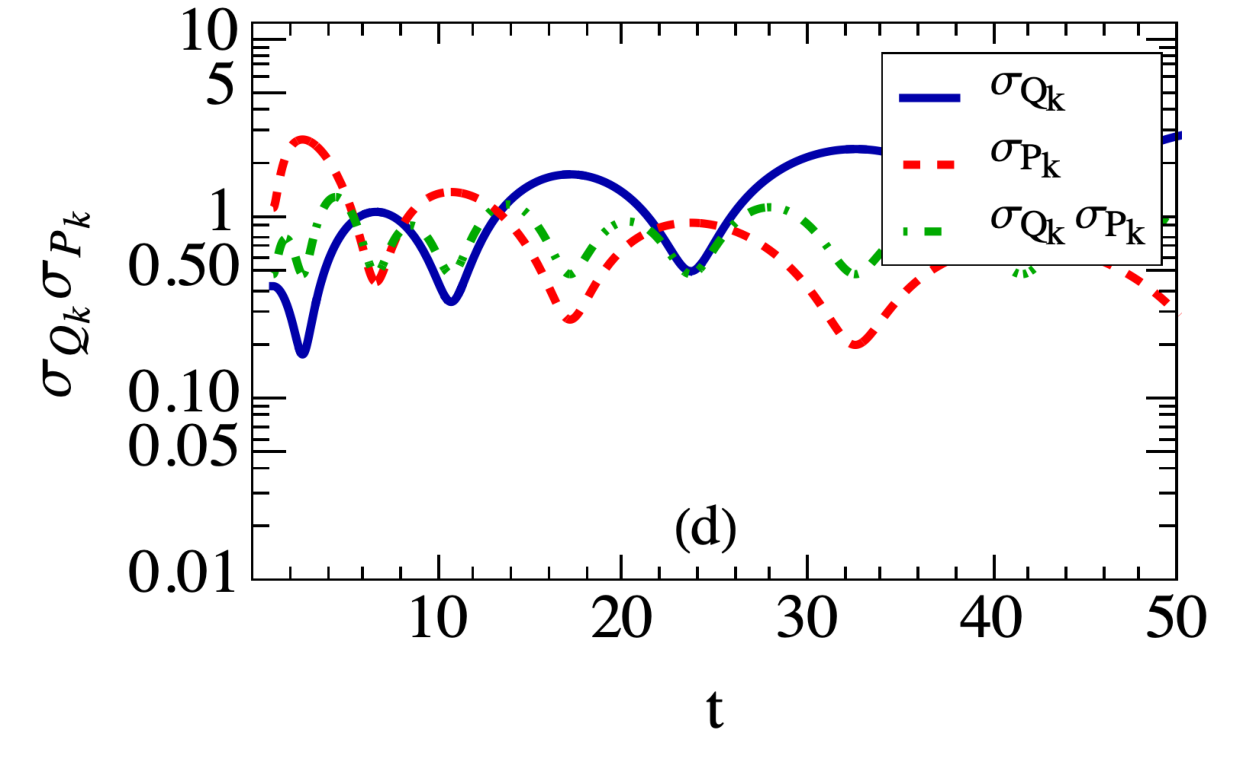}}\caption{A panel of the uncertainty relation plots which consider the radiation
era $\left(\alpha=2\right)$. Figs. (a) and (b) we set $r=0.5$, while
$L=10^{-2}$ and $L=1$, respectively. Figs. (c) and (d) we have fixed
$r=-0.5$, with $L=10^{-2}$ and $L=1$, respectively.}
\label{fig1} 
\end{figure}

\begin{figure}[ptb]
\subfloat{\includegraphics[scale=0.25]{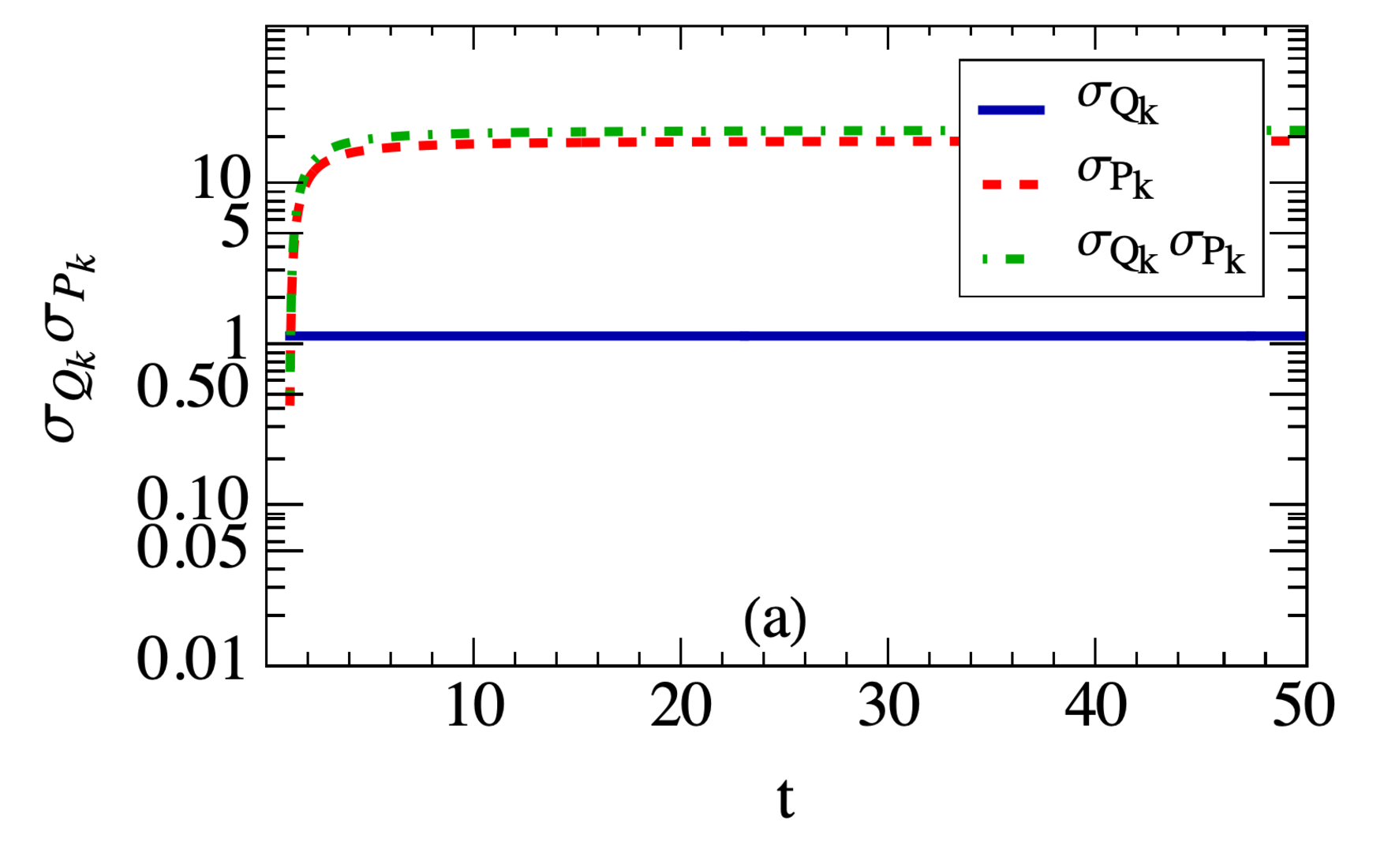}}\hfill{}\subfloat{\includegraphics[scale=0.25]{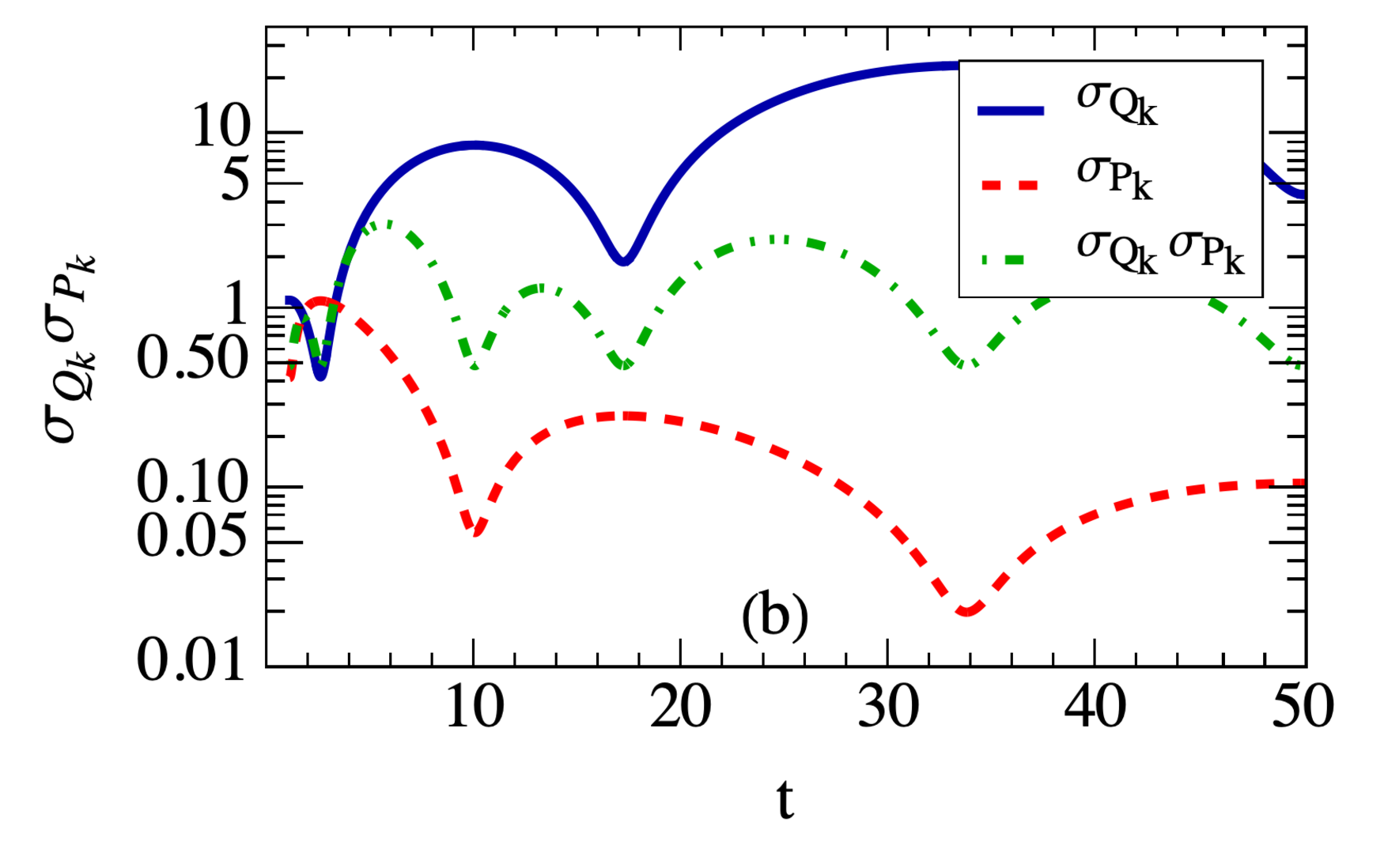}}\vfill{}
 \subfloat{\includegraphics[scale=0.25]{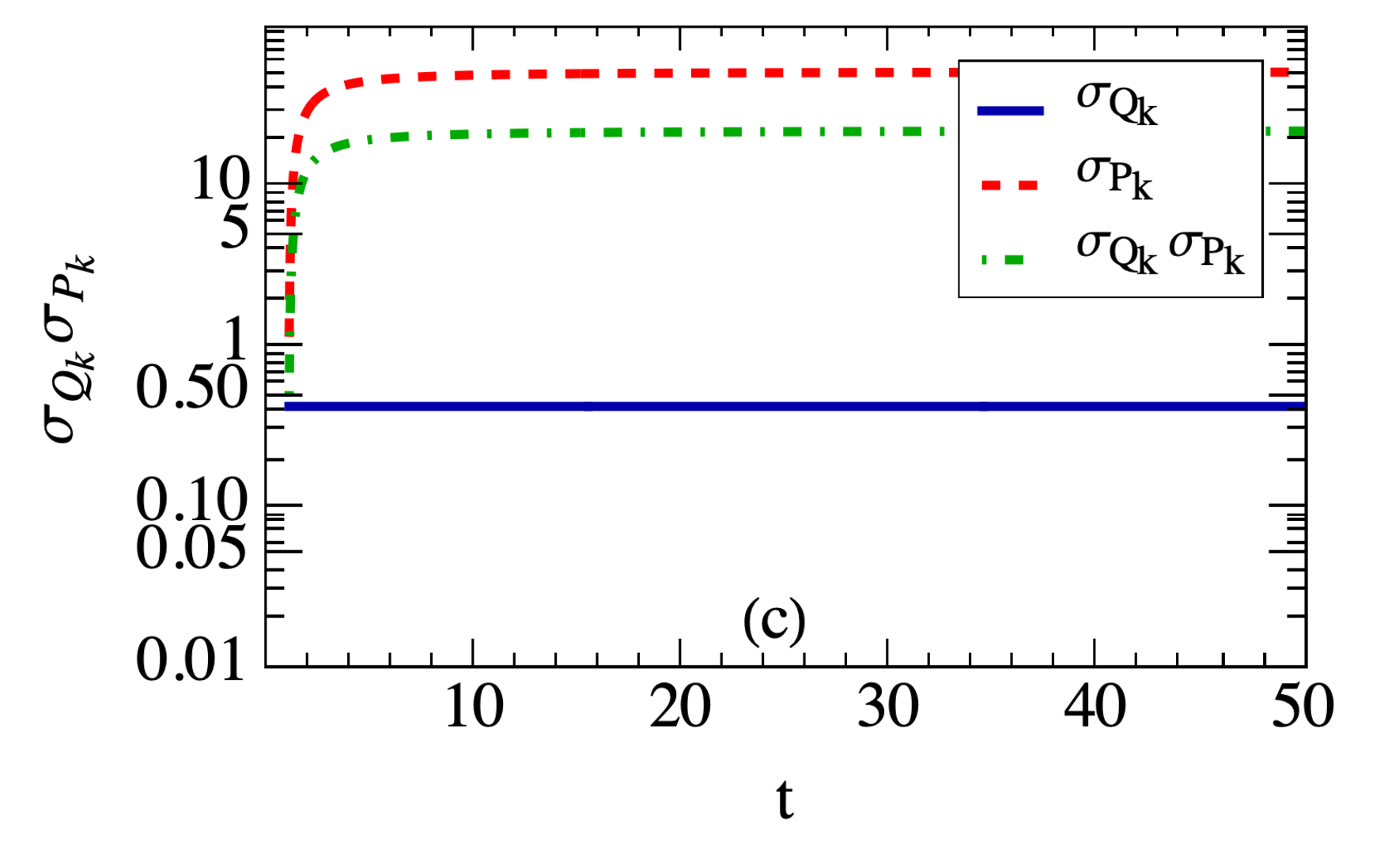}}\hfill{}\subfloat{\includegraphics[scale=0.25]{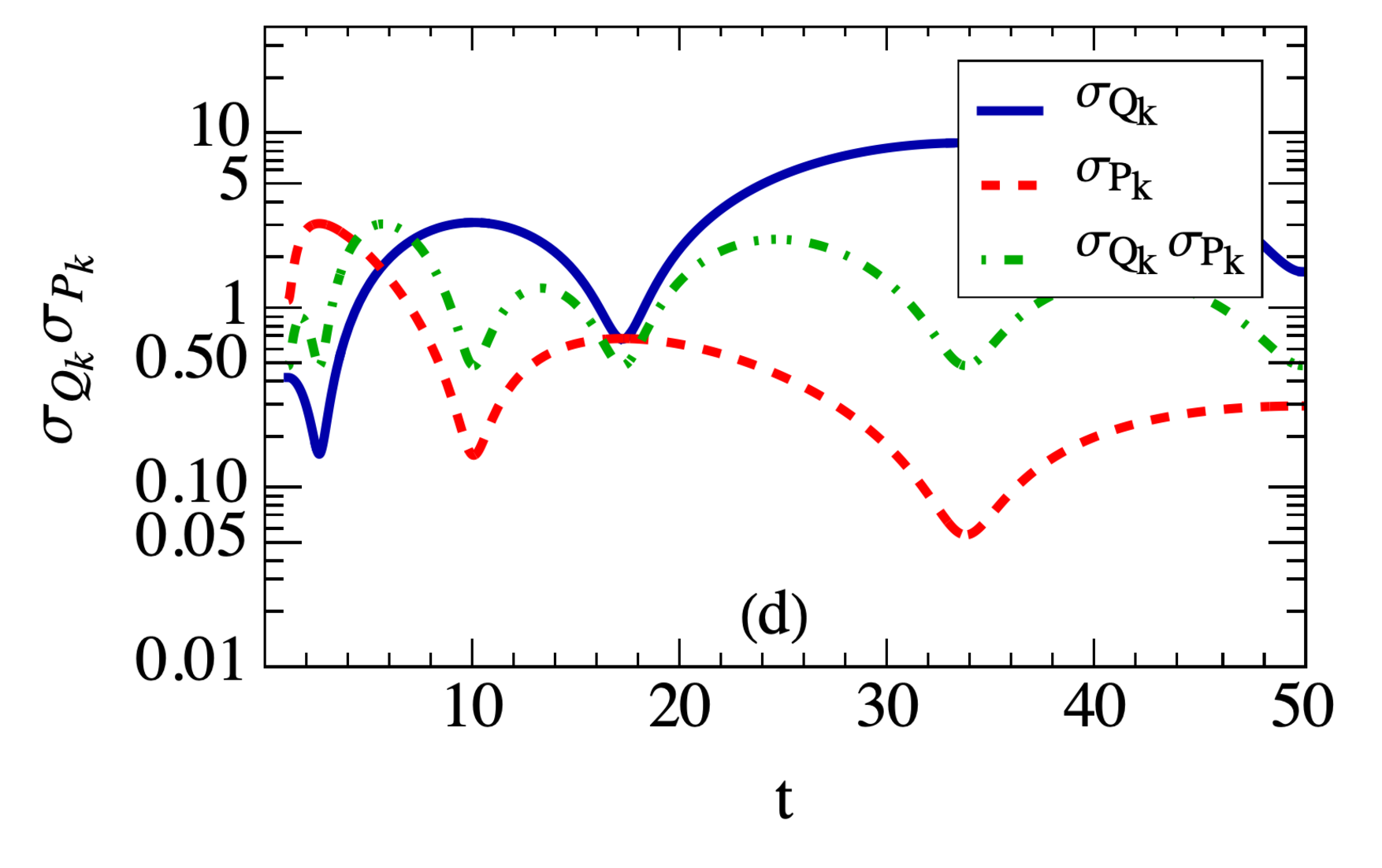}}\caption{Uncertainty relations for $\alpha=3/2$ (matter era). Figs. (a) and
(b) we assume $r=0.5$, while $L=10^{-2}$ and $L=1$, respectively.
Figs. (c) $L=10^{-2}$ and (d) $L=1$ we consider $r=-0.5$.}
\label{fig2} 
\end{figure}

We should highlight that the uncertainties on the field and momentum
depend on the extra coordinate, $L$, and the $\alpha$-parameter
that rules the cosmological eras under consideration. In this case,
in Figs. \ref{fig1} and \ref{fig2}, we present an investigation
of the space of parameters considering the previously uncertainties
here derived {[}see Eq. (\ref{M12}){]}. Therefore, as one can see,
the $r$-parameter is responsible for magnifying the squeeze properties
on both quadratures. The Figures \ref{fig1}(a) and \ref{fig1}(c)
unveil that negative values of $r$ squeeze $\sigma_{P_{k}}$ standard
deviation, while positive values squeeze $\sigma_{Q_{k}}$. It is
possible to verify that the variances for the SCSs are lower than
the generalized uncertainty minimum, confirming the non-classicality
of constructed states. On the other hand, the extra dimension $L$
endows oscillatory pattern to the uncertainties, which could, in principle,
represent a signature of the extra dimension, as can be seen in Figs.
\ref{fig1} and \ref{fig2} {[}(b) and (d){]}. Finally, the $\alpha$-parameter,
which identifies the cosmological era, gives rise to a bouncing behavior
for the uncertainty relations. So, in the matter era $\left(\alpha=3/2\right)$,
we observe a slow oscillation in the course its time evolution, compared
to the radiation-dominated era $\left(\alpha=2\right)$. We conclude
this section by stressing that the inherent characteristic of the
SCSs, known as the squeezing property, has been demonstrated from
this analysis.

\subsection{Representation of the SCSs in terms of the field $\phi_{k}$}

Here, we intend to present the SCSs in another physical representation.
With this aim, we must follow some steps. First, we find the vacuum
state $\Psi_{0_{k}}\left(\phi_{k}\right)\equiv\left\langle \phi_{k}|0_{k}\right\rangle $
through the condition of annihilation $\hat{a}_{k}\Psi_{0_{k}}\left(\phi_{k}\right)=0$
so that we use the normalization condition, such that, 
\begin{align}
 & \left(\frac{\phi_{k}}{l_{k}^{2}}+\frac{\partial}{\partial\phi_{k}}\right)\Psi_{0_{k}}\left(\phi_{k}\right)=0,\nonumber \\
 & \Psi_{0_{k}}\left(\phi_{k}\right)=\frac{1}{\sqrt{l_{k}\sqrt{\pi}}}\exp\left(-\frac{\phi_{k}^{2}}{2l_{k}^{2}}\right),\text{ \ }\int_{-\infty}^{\infty}\Psi_{0_{k}}^{2}\left(\phi_{k}\right)d\phi_{k}=1.\label{R1}
\end{align}
Therefore, by applying $n$ times the creation operator $\hat{a}_{k}^{\dagger}$
to $\Psi_{0}\left(\phi_{k}\right)$, we find the following form for
the $\Psi_{n}\left(\phi_{k}\right)$ states 
\begin{equation}
\Psi_{n_{k}}\left(\phi_{k}\right)=\frac{\left(\hat{a}_{k}^{\dagger}\right)^{n_{k}}}{\sqrt{n_{k}!}}\Psi_{0_{k}}\left(\phi_{k}\right)=\frac{H_{n_{k}}\left(\frac{\phi_{k}}{l_{k}}\right)}{\sqrt{l_{k}\sqrt{\pi}2^{n_{k}}n_{k}!}}\exp\left(-\frac{\phi_{k}^{2}}{2l_{k}^{2}}\right).\label{R2}
\end{equation}
Finally, from (\ref{W11}) and (\ref{R2}), we have the representation
of SCSs in terms of fields $\phi_{k}$, $\Psi_{\zeta_{k},\xi_{k}}\left(\phi_{k}\right)\equiv\left\langle \phi_{k}|\zeta_{k},\xi_{k}\right\rangle $,
which is explicitly written as 
\begin{align}
 & \Psi_{\zeta_{k},\xi_{k}}\left(\phi_{k},t\right)=\frac{\left(1-\left\vert \zeta_{k}\right\vert ^{2}\right)^{1/4}}{\sqrt{\sqrt{2\pi}e^{r}\sigma_{\phi_{0k}}\left(1-\zeta_{k}\right)}}\exp\left[-\frac{1+\zeta_{k}}{4e^{2r}\sigma_{\phi_{0k}}^{2}\left(1-\zeta_{k}\right)}\left(\phi_{k}+\frac{2e^{r}\sigma_{\phi_{0k}}\xi_{k}}{1+\zeta_{k}}\right)^{2}\right.\nonumber \\
 & \left.+\frac{\left(1+\zeta_{k}^{\ast}\right)\xi_{k}^{2}}{2\left(1+\zeta_{k}\right)\left(1-\left\vert \zeta_{k}\right\vert ^{2}\right)}-\frac{\left\vert \xi_{k}\right\vert ^{2}}{2\left(1-\left\vert \zeta_{k}\right\vert ^{2}\right)}+i\theta_{k}\right].\label{R3}
\end{align}
On the other hand, in terms of the $\overline{\phi_{k}}$ and $\overline{\Pi_{k}}$
means values, we have 
\begin{equation}
\Psi_{\overline{\phi_{k}}}^{\sigma_{\phi_{k}}}\left(\phi_{k},t\right)=\frac{e^{i\vartheta_{k}}}{\sqrt{\sqrt{2\pi}\sigma_{\phi_{k}}}}\exp\left[-\frac{\left(\phi_{k}-\overline{\phi_{k}}\right)^{2}}{4\sigma_{\phi_{k}}^{2}}\right],\label{R4}
\end{equation}
where 
\begin{align}
 & e^{i\vartheta_{k}}=\left(\frac{1-\zeta_{k}^{\ast}}{1-\zeta_{k}}\right)^{1/4}\exp\left[\frac{i\sigma_{\phi_{k}\Pi_{k}}}{\hbar\sigma_{\phi_{k}}^{2}}\frac{\left(\phi_{k}-\overline{\phi_{k}}\right)^{2}}{2}+\frac{i\overline{\Pi_{k}}}{2\hbar}\left(2\phi_{k}-\overline{\phi_{k}}\right)+i\theta_{k}\right],\nonumber \\
 & \overline{\phi_{k}}=-\frac{2e^{r}\sigma_{\phi_{0k}}\operatorname{Re}\left[\left(1-\zeta_{k}^{\ast}\right)\xi_{k}\right]}{1-\left\vert \zeta_{k}\right\vert ^{2}},\text{ \ }\sigma_{\phi_{k}}=e^{r}\sigma_{\phi_{0k}}\frac{\left\vert 1-\zeta_{k}\right\vert }{\sqrt{1-\left\vert \zeta_{k}\right\vert ^{2}}},\nonumber \\
 & \xi_{k}=\frac{4\hbar L_{k}\left(\tau t\right)^{\frac{\varsigma-\delta}{2}}\varphi_{k}}{\pi\hbar\beta_{k}L_{k}\left[\tau^{\varsigma}W_{k,\varepsilon,\varepsilon-1}^{\left(\tau,t\right)}e^{r}+t^{\varsigma}W_{k,\varepsilon,\varepsilon-1}^{\left(t,\tau\right)}e^{-r}\right]+i\pi\beta_{k}\left[\hbar^{2}\left(\tau t\right)^{\varsigma}W_{k,\varepsilon,\varepsilon}^{\left(\tau,t\right)}e^{r}-L_{k}^{2}W_{k,\varepsilon-1,\varepsilon-1}^{\left(t,\tau\right)}e^{-r}\right]},\nonumber \\
 & \zeta_{k}=\frac{\hbar L_{k}\left[\tau^{\varsigma}W_{k,\varepsilon,\varepsilon-1}^{\left(\tau,t\right)}e^{r}-t^{\varsigma}W_{k,\varepsilon,\varepsilon-1}^{\left(t,\tau\right)}e^{-r}\right]-i\left[\hbar^{2}\left(\tau t\right)^{\varsigma}W_{k,\varepsilon,\varepsilon}^{\left(\tau,t\right)}e^{r}+L_{k}^{2}W_{k,\varepsilon-1,\varepsilon-1}^{\left(t,\tau\right)}e^{-r}\right]}{\hbar L_{k}\left[\tau^{\varsigma}W_{k,\varepsilon,\varepsilon-1}^{\left(\tau,t\right)}e^{r}+t^{\varsigma}W_{k,\varepsilon,\varepsilon-1}^{\left(t,\tau\right)}e^{-r}\right]+i\left[\hbar^{2}\left(\tau t\right)^{\varsigma}W_{k,\varepsilon,\varepsilon}^{\left(\tau,t\right)}e^{r}-L_{k}^{2}W_{k,\varepsilon-1,\varepsilon-1}^{\left(t,\tau\right)}e^{-r}\right]},\nonumber \\
 & \varsigma\equiv\delta\left(2\varepsilon-1\right),\text{ \ }L_{k}=\frac{2\delta m_{0}\beta_{k}e^{2r}}{\tau^{\frac{3+\alpha}{\alpha}}}\sigma_{\phi_{0k}}^{2}.\label{R5}
\end{align}

The representation found in Eq. (\ref{R4}) will allow us to discuss
the probability density, about which we will investigate the influence
of both the extra dimension $L$ and the cosmological era from the
parameter $\alpha$.

\subsubsection{Probability density}

\begin{figure}[ptb]
\subfloat{\includegraphics[scale=0.33]{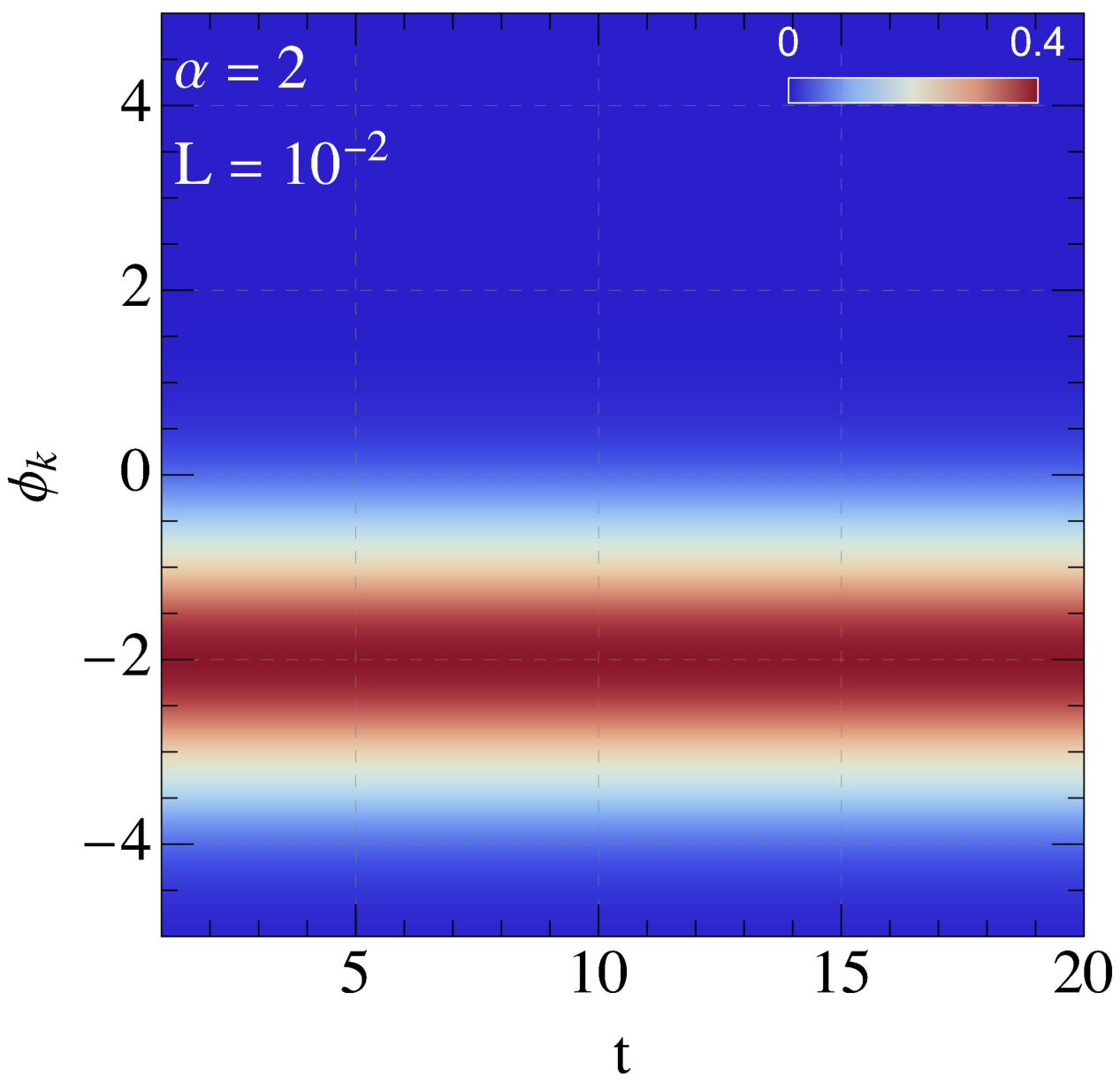}}\,\subfloat{\includegraphics[scale=0.33]{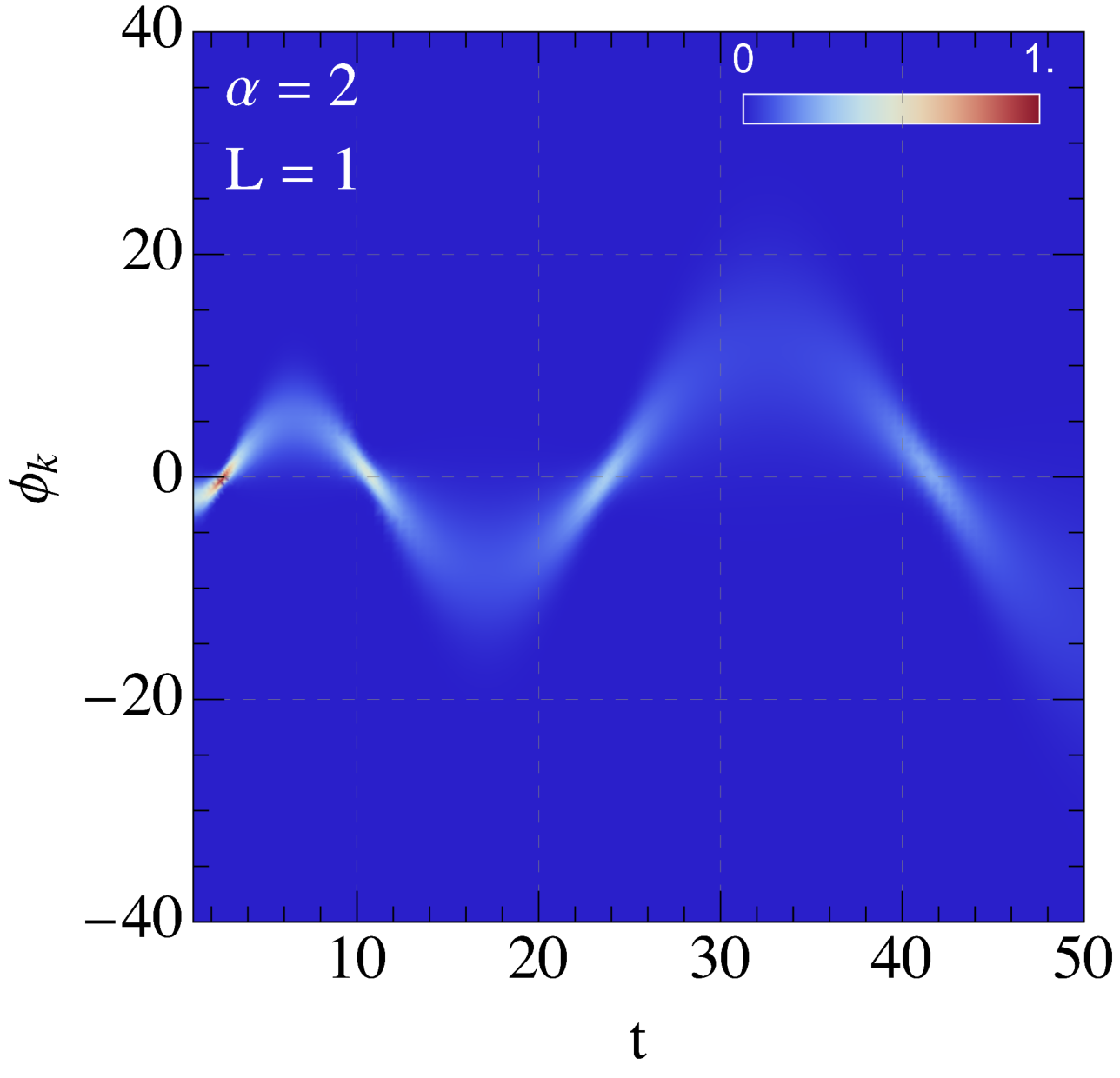}}\vfill{}
 \subfloat{\includegraphics[scale=0.33]{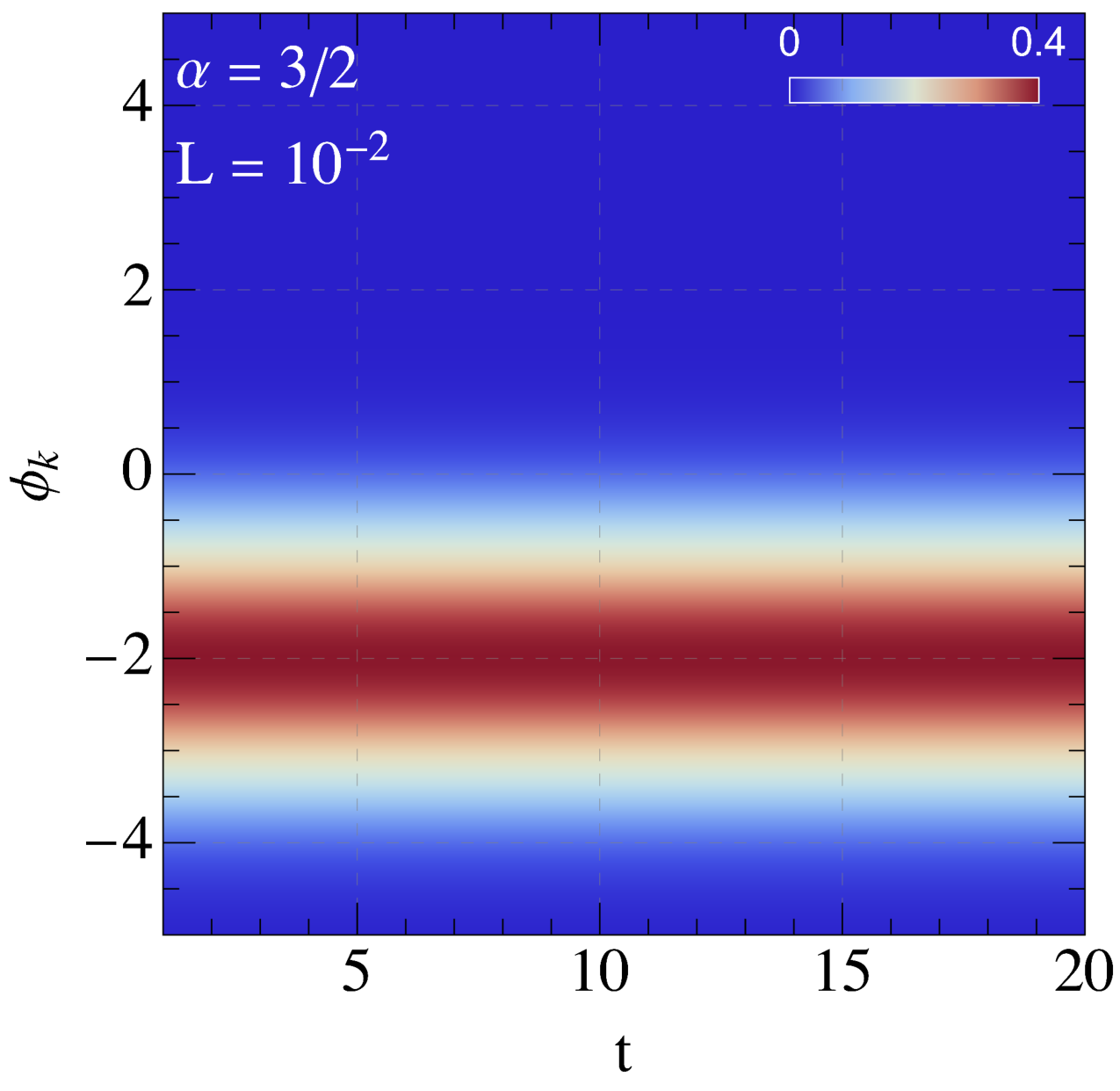}}\,\subfloat{\includegraphics[scale=0.33]{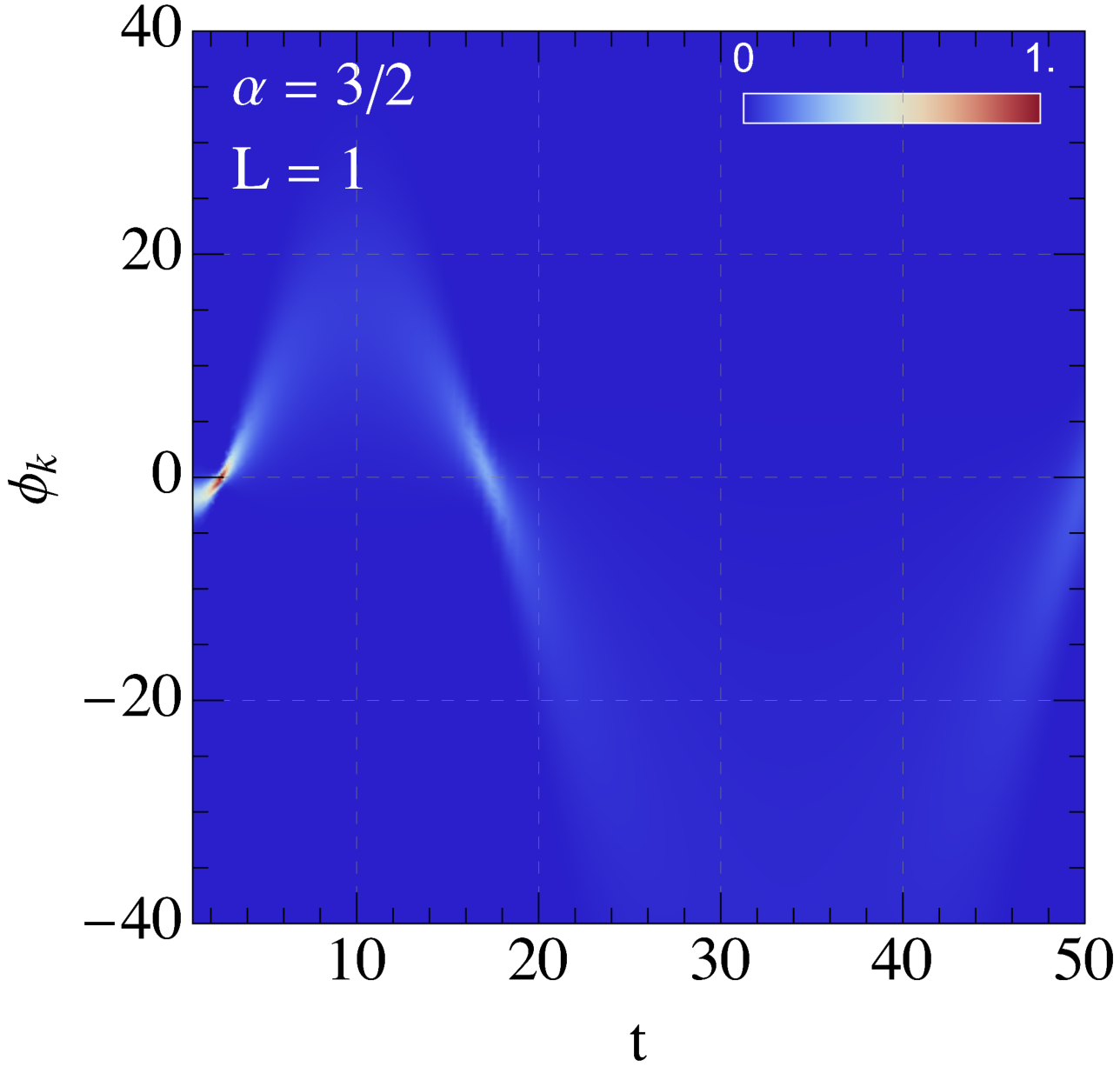}}\caption{Probability density contour plots $\left(t,\phi_{k}\right)$ for the
inflaton-like field considering the radiation era ($\alpha=2$), and
matter era ($\alpha=3/2$). }
\label{fig3} 
\end{figure}

Here we aim to study the probability density, which is defined as
$\rho_{\overline{\phi_{k}}}^{\sigma_{\phi_{k}}}\left(\phi_{k},t\right)\equiv\left\vert \Psi_{\overline{\phi_{k}}}^{\sigma_{\phi_{k}}}\left(\phi_{k},t\right)\right\vert ^{2}$.
Thus, from Eq. (\ref{R4}), we can readily obtain 
\begin{equation}
\rho_{\overline{\phi_{k}}}^{\sigma_{\phi_{k}}}\left(\phi_{k},t\right)=\frac{1}{\sqrt{2\pi}\sigma_{\phi_{k}}}\exp\left[-\frac{\left(\phi_{k}-\overline{\phi_{k}}\right)^{2}}{2\sigma_{\phi_{k}}^{2}}\right],\label{R6-1}
\end{equation}
which is called the Gaussian or normal distribution and has been shown
in Figure \ref{fig3}. To depict these Figures we have assumed $\hbar=1,\varphi_{k}=1,\sigma_{\phi_{0k}}=1,k=1,\tau=1$,
and $r=0.5$. As we see, once again, the extra dimension, $L$, is
responsible for the emergence of oscillatory behavior on the density,
whereas the cosmological parameter $\alpha$ will be related to the
amplitude and frequency of oscillations. Besides, small values for
the $L$-parameter result in a constant probability density in the
time, while large values change significantly its behavior, showing,
in this case, an SCSs peak for initial times ($t\sim2$). In this
case, it seems this physical system can be employed to seek traces
of extra dimensions.

\section{Particle creation in SCSs\label{sec5}}
The 5D Kaluza-Klein-type cosmological models described
by metric (\ref{metric}) are known to contain 4D hypersurfaces that accurately
mimic the standard FLRW cosmologies, besides featuring the Big Bang
singularity \citep{Fukui2001}. Indeed, the density and pressure for
the metric (\ref{metric}) is given by:
\begin{equation}
\rho=\frac{3}{\alpha^{2}\ell^{2}t^{2}},\qquad p=\frac{\left(2\alpha-3\right)}{\alpha^{2}\ell^{2}t^{2}}.
\end{equation}These results leads to the equation of state $p=(2\alpha/3-1)\rho$. Therefore, if we consider $\alpha=3/2$ and $\ell=1$,
we recover the usual $k=0$ model, which describes the late Universe. On the other hand, for $\alpha=1/3$ and $\ell\gg1$, we obtain the
inflationary $k=0$ models. Consequently, the properties of matter
are identical in both five and four dimensions \citep{Wesson1992}.
Now, by assuming a transformation so that the proper
time is defined by $T=\ell t$, the Hubble parameter takes the
form
\begin{equation}
H=\frac{1}{\alpha T}.
\end{equation}
Thus, we can see that the 5D solution (\ref{metric}) encompasses 4D dynamics
and matter that match those of standard 4D cosmologies for both the
early and late Universe \citep{Wesson1995}. We must, therefore, circumvent
this apparent impossibility of seeking traces of extra dimensions
via analysis of the cosmological evolution of the Universe. 

Occurs that, quite generically, the coupling of a scalar field to a curved background,
in general, yields to cosmological particle production \citep{Parker1968,Parker1969,Parker1971,Parker2009}.
On its turn, the emergence of squeezed quantum states is an unavoidable
consequence whenever particle creation takes place \citep{Grishchuk1990}.
Therefore, in this section we aimed to investigate the particle creation
of the inflaton-like field in a Kaluza-Klein Universe compatible with
the SSs formalism from quantum optics. In this case, the number of
particles at a later time, $t$, created from the vacuum at the initial
time, $\tau$, is represented by the following equation: 
\begin{equation}
\overline{n}_{k}=\left\langle \zeta_{k},\xi_{k}\left\vert \hat{n}_{k}\right\vert \xi_{k},\zeta_{k}\right\rangle =\frac{\left\vert \zeta_{k}\right\vert ^{2}\left(1-\left\vert \zeta_{k}\right\vert ^{2}\right)+\left\vert \zeta_{k}\xi_{k}^{\ast}-\xi_{k}\right\vert ^{2}}{\left(1-\left\vert \zeta_{k}\right\vert ^{2}\right)^{2}},\label{number}
\end{equation}
where $\hat{n}_{k}=\hat{a}^{\dagger}\hat{a}$ is the number operator,
and the state $\left|\xi_{k},\zeta_{k}\right\rangle $ is given by
(\ref{W11}). The Eq. (\ref{number}) represents the average number
of particles in the $k$-mode at time $t$.

\begin{figure}[ptb]
\subfloat{\includegraphics[scale=0.2]{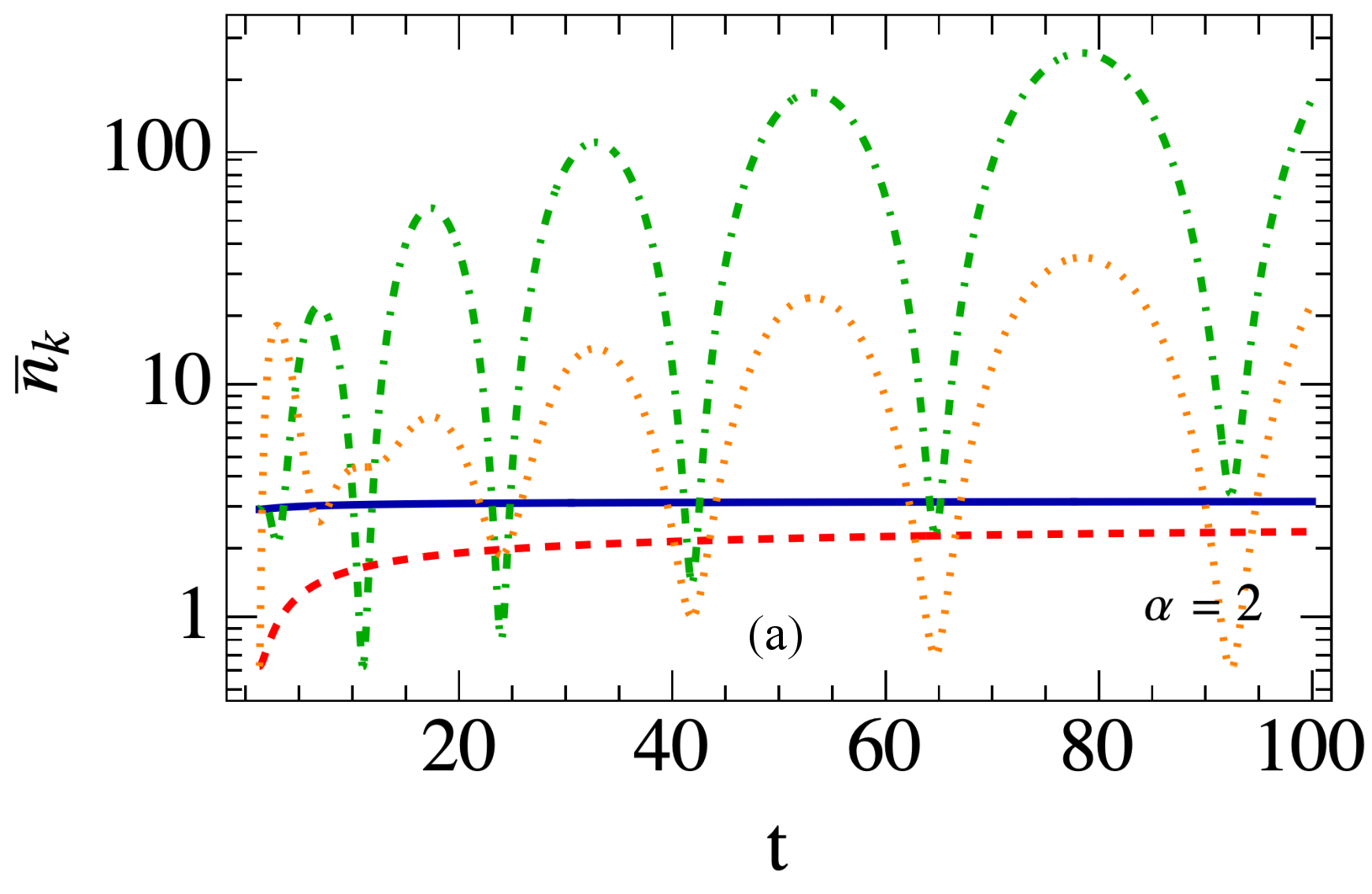}}\,\subfloat{\includegraphics[scale=0.2]{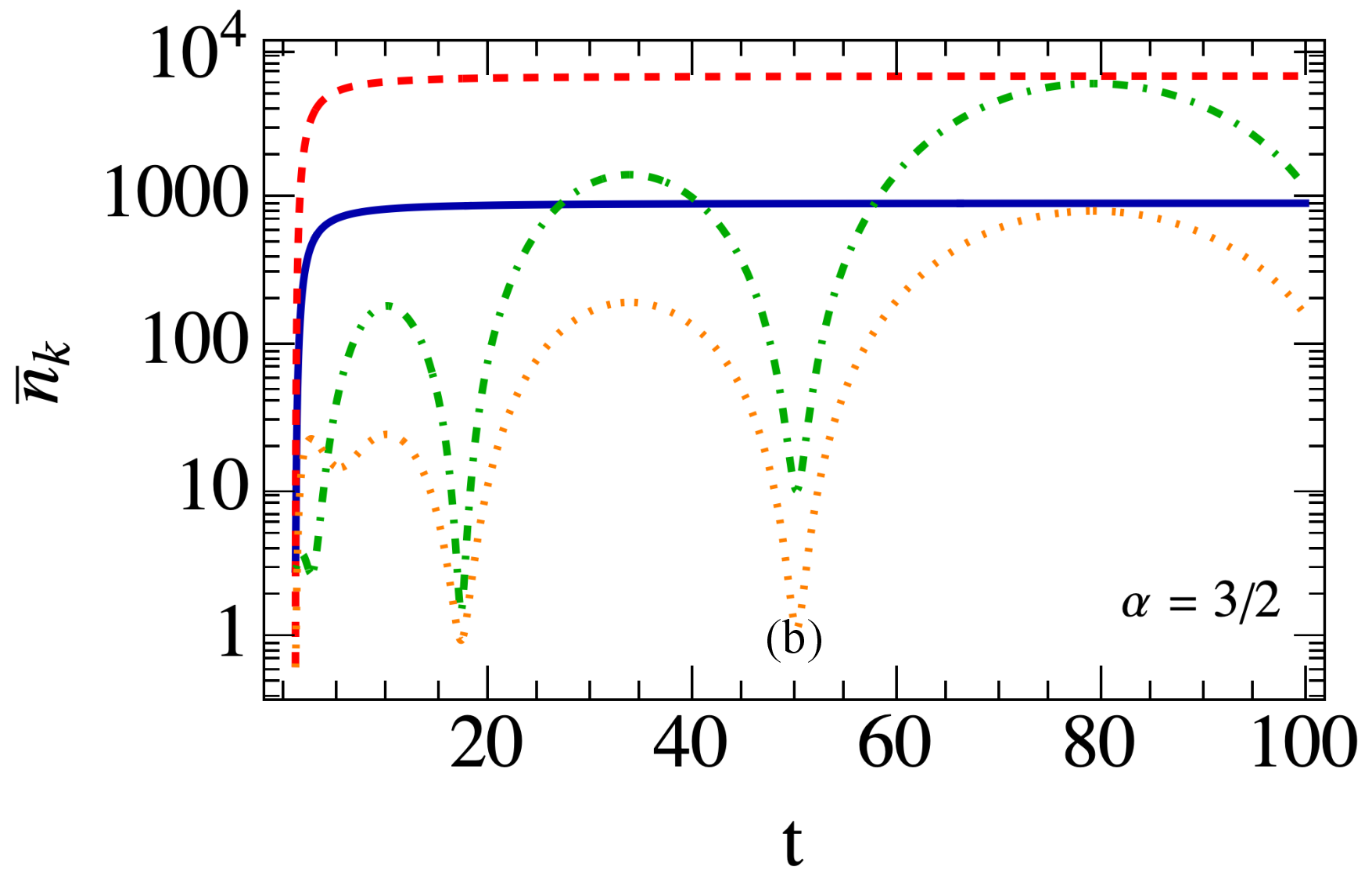}}\,\subfloat{\includegraphics[scale=0.2]{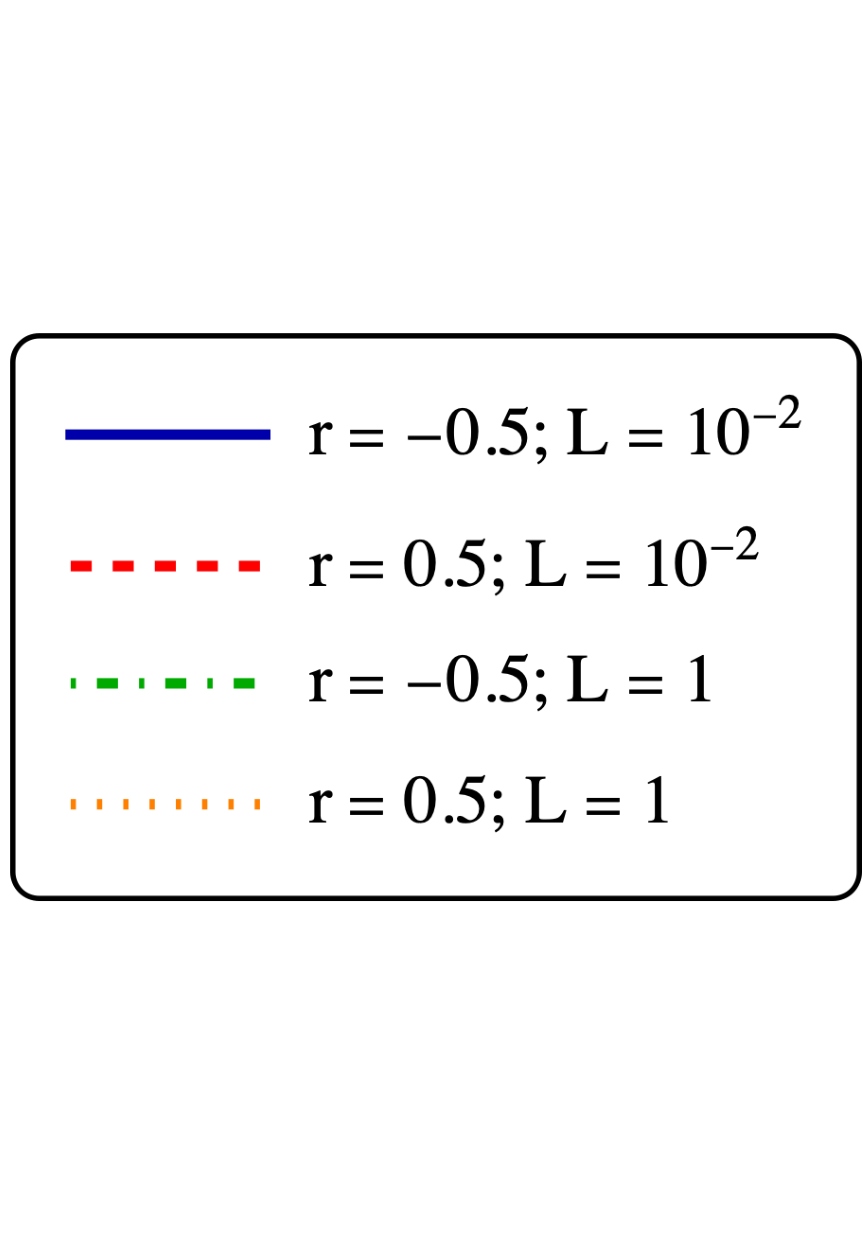}}\caption{Number of created particles in SCSs are shown for the case of a (a)
radiation and (b) matter eras of the Universe.}
\label{fig4} 
\end{figure}

As we have shown in Figure \ref{fig4}, the number of created particles
$\overline{n}_{k}$ is finite and will grow to a constant maximum
value in time if the extra dimension $L\ll1$. In this case, the displacement-
and $r$-parameters will be responsible to control this limit and,
therefore, are related to the excitation of the system. On the other
hand, if $L\sim1$, we note that the number of created particles oscillates
with time. Thus, the extra dimension leaves a clear fingerprint on
this physical observable, and therefore, in principle, this physical
system can be eventually applied to impose constraints on the extra
dimension. In its turn, if $L\ll1$, we find that the number of produced
particles in SCSs for the matter era is greater than the one found
in the radiation era.

Furthermore, we must highlight that Eq. (\ref{number}) describes
the number of particles produced in the coherent states (CSs) for
$\zeta_{k}=0$. This condition can be determined from Eq. (\ref{Q8}).
Note that, in this case, we should have $g_{k}=0$. However, such
an imposition would imply that $f_{k}=0$ or $\eta_{-}=0$. Then,
in order to avoid the trivial solution, we must consider 
\begin{equation}
\eta_{-}=0\rightarrow t_{0}=\left(\frac{\hbar\left|\alpha-1\right|}{\alpha kl_{k}^{2}L^{(\alpha-5)/2(\alpha-1)}}\right)^{\alpha/\left(\alpha+2\right)}.
\end{equation}
As we see, inflaton particles in CSs only are allowed at the initial
time, i.e., the time for which the uncertainty relation is minimized.
Once that, in this case, the number of particles depends on the extra
dimension and on the cosmological parameter. The constraints over
the inflaton field in CSs also may be a promising way to search for
signatures of extra dimensions. On the other hand, for $\xi_{k}=0$
($\varphi_{k}=0$), Eq. (\ref{number}) describes the number of particles
produced in the SSs.

\section{Final Remarks\label{sec6}}

In this work, we study a propagating massless scalar field on the
Kaluza-Klein-type cosmological background. In this context, by investigating
the Hamiltonian associated with the framework described in addition
to the respective equations of motion, one can found that this physical
system resembles a damped oscillator with both time-dependent mass
and frequency. Besides, we show that the $5D$ massless field can
be identified with a massive $4D$ inflaton field.

Following the SSs formalism of quantum mechanics, we have constructed
the SCSs for the quantized field in a non-unitary approach considering
the invariant operator method of Lewis-Riesenfeld. The uncertainty
relations have been investigated, and from a broad study of the parameter
space, their minimization was analyzed. Moreover, we have shown that
the obtained states have the squeezing property. We found that the
extra dimension, $L$, is accountable for providing an oscillatory
behavior presented on the uncertainty relations, while the cosmological
eras, controlled by the $\alpha$-parameter, regulates the frequency
of these oscillations. In this case, for large values of the extra
dimension, the oscillations on uncertainty relations give rise. This
seems in total agreement with the fact that $5D$ massless inflaton
wave function can access our $4D$ Universe as we increase the size
of the extra-dimension. The probability density and transition probability
have been calculated, and again the extra dimension accounts for raising
the oscillatory behavior on these quantities. Moreover, since the
cosmological expansion can generate particles, we investigated particle
creation in SCSs in Kaluza-Klein cosmological scenario. By considering
a small value for the extra dimension, i.e., $L\ll1$, the number
of created particles in SCSs reaches a maximum value. Besides, if
we work with higher values for this parameter, i.e., $L\sim1$, the
particle production bounces in time. Finally, it is worth mentioning
that once the extra dimension prints an unmistakable signature to
all quantities studied, perhaps this is a promising alternative physical
system to seek traces of hidden dimensions.

The methodology here presented can be implemented to several other
geometrical backgrounds, such as braneworld scenarios \citep{Clifton2012}.
Moreover, these discussions applied to the analogue gravity models
\citep{Barcelo2005} can provide crucial insights into the Universe's
structure. Shortly, we expect to report on some of these topics. 
\begin{acknowledgments}
We would like to thank CNPq, CAPES and CNPq/PRONEX/FAPESQ-PB (Grant
No. 165/2018), for partial financial support. F.~A.~B. acknowledges support
from CNPq (Grant No. 309092/2022-1). A.~S.~L. acknowledges support from
CAPES (Grant No. 88887.800922/2023-00). J.~R.~L.~S. acknowledges CNPq (Grant
No. 309494/2021-4), and PRONEX/CNPq/FAPESQ-PB (Grant Nos. 165/2018,
and 0015/2019) for financial support. A.~S.~P. thanks the support of the
Instituto Federal da Para\'{i}ba. 
\end{acknowledgments}

\appendix
%dummy comment inserted by tex2lyx to ensure that this paragraph is not empty%dummy comment inserted by tex2lyx to ensure that this paragraph is not empty

\section{Solutions for $f_{k}$ and $g_{k}$ \label{Appendix-A}}

Let us start considering the differential equations 
\begin{equation}
\dot{f}_{k}=i\left(\eta_{+}f_{k}-\eta_{-}g_{k}\right),\text{ \ }\dot{g}_{k}=i\left(\eta_{-}f_{k}-\eta_{+}g_{k}\right).\label{B1}
\end{equation}
Now, by defining the new quantities 
\begin{equation}
F_{k}=f_{k}+g_{k}\Rightarrow\dot{F}_{k}=\dot{f}_{k}+\dot{g}_{k},\text{ \ }G_{k}=f_{k}-g_{k}\Rightarrow\dot{G}_{k}=\dot{f}_{k}-\dot{g}_{k},\label{B2}
\end{equation}
one may rewrite the equations (\ref{B1}) in the form 
\begin{equation}
\dot{F}_{k}=i\left(\eta_{+}+\eta_{-}\right)G_{k},\text{ \ }\dot{G}_{k}=i\left(\eta_{+}-\eta_{-}\right)F_{k}.\label{B3}
\end{equation}
On the other hand, knowing that 
\begin{align}
\eta_{+}+\eta_{-} & =\frac{l_{k}^{2}m\omega_{k}^{2}}{\hbar}\Rightarrow\dot{\eta}_{+}+\dot{\eta}_{-}=\frac{1+\alpha}{\alpha t}\left(\eta_{+}+\eta_{-}\right),\nonumber \\
\eta_{+}-\eta_{-} & =\frac{\hbar}{l_{k}^{2}m}\Rightarrow\dot{\eta}_{+}-\dot{\eta}_{-}=-\frac{3+\alpha}{\alpha t}\left(\eta_{+}-\eta_{-}\right),\label{B4}
\end{align}
we get, 
\begin{equation}
\ddot{G}_{k}=i\left(\dot{\eta}_{+}-\dot{\eta}_{-}\right)F_{k}+i\left(\eta_{+}-\eta_{-}\right)\dot{F}_{k}=-\frac{3+\alpha}{\alpha t}\dot{G}_{k}-\left(\eta_{+}^{2}-\eta_{-}^{2}\right)G_{k}.\label{B5}
\end{equation}
Furthermore, one can check that $\eta_{+}^{2}-\eta_{-}^{2}=\omega_{k}^{2}$.
Thus, we can rewritten (\ref{B5}) as follows 
\begin{equation}
\ddot{G}_{k}+\frac{3+\alpha}{\alpha t}\dot{G}_{k}+\omega_{k}^{2}G_{k}=0.\label{B7}
\end{equation}
From the Eq. (\ref{B3}), we find $F_{k}$, in the form 
\begin{equation}
F_{k}=-\frac{i}{\eta_{+}-\eta_{-}}\dot{G}_{k}=-\frac{il_{k}^{2}m}{\hbar}\dot{G}_{k}.\label{B8}
\end{equation}

Considering equations (\ref{B7}) and (\ref{B8}) with initial conditions
$F_{0k}=F_{k}\left(\tau\right)$ and $G_{0k}=G_{k}\left(\tau\right)$,
we will obtain 
\begin{align}
 & G_{k}=\frac{\pi\beta_{k}t^{\delta\varepsilon}}{2\tau^{\delta\left(\varepsilon-1\right)}}W_{k,\varepsilon,\varepsilon-1}^{\left(t,\tau\right)}G_{0k}+\frac{i\pi\hbar\left(\tau t\right)^{\delta\varepsilon+\frac{3+\alpha}{\alpha}}}{2l_{k}^{2}\delta m}W_{k,\varepsilon,\varepsilon}^{\left(\tau,t\right)}F_{0k},\nonumber \\
 & F_{k}=\frac{\pi\beta_{k}\tau^{\delta\varepsilon}}{2t^{\delta\left(\varepsilon-1\right)}}W_{k,\varepsilon,\varepsilon-1}^{\left(\tau,t\right)}F_{0k}-\frac{i\pi l_{k}^{2}\delta\beta_{k}^{2}m}{2\hbar\left(\tau t\right)^{\delta\varepsilon+\frac{4}{\alpha}}}W_{k,\varepsilon-1,\varepsilon-1}^{\left(t,\tau\right)}G_{0k},\label{B9}
\end{align}
where $\tau$ is the initial time and 
\begin{align}
 & \varepsilon=\frac{3}{2\left\vert \alpha-1\right\vert },\text{ \ }\beta_{k}=\frac{\alpha\omega_{0k}\tau^{1/\alpha}}{\left\vert \alpha-1\right\vert },\text{ \ }\delta=\frac{\alpha-1}{\alpha},\text{ \ }W_{k,a,b}^{\left(\tau,t\right)}=J_{a}\left(\beta_{k}\tau^{\delta}\right)Y_{b}\left(\beta_{k}t^{\delta}\right)-J_{b}\left(\beta_{k}t^{\delta}\right)Y_{a}\left(\beta_{k}\tau^{\delta}\right),\nonumber \\
 & W_{k,\varepsilon,\varepsilon-1}^{\left(\tau,\tau\right)}=\frac{2}{\pi\beta_{k}\tau^{\delta}},\text{ \ }W_{k,b,a}^{\left(\tau,t\right)}=-W_{k,a,b}^{\left(t,\tau\right)},\text{ \ }W_{k,a,a}^{\left(\tau,\tau\right)}=0.\label{B9a}
\end{align}

Finally, the functions $f_{k}$ and $g_{k}$ are determined from the
relations below 
\begin{equation}
f_{k}=\frac{F_{k}+G_{k}}{2},\text{ \ }g_{k}=\frac{F_{k}-G_{k}}{2}.\label{B10}
\end{equation}
Thus, for the $f_{k}$- and $g_{k}$-functions, we get 
\begin{align}
f_{k}= & \frac{\pi\beta_{k}\tau^{\delta}}{4}\left[\frac{\tau^{\delta\left(\varepsilon-1\right)}}{t^{\delta\left(\varepsilon-1\right)}}W_{k,\varepsilon,\varepsilon-1}^{\left(\tau,t\right)}F_{0k}+\frac{t^{\delta\varepsilon}}{\tau^{\delta\varepsilon}}W_{k,\varepsilon,\varepsilon-1}^{\left(t,\tau\right)}G_{0k}\right]\nonumber \\
+ & \frac{i\pi\hbar\left(\tau t\right)^{\delta\varepsilon+\frac{3+\alpha}{\alpha}}}{4l_{k}^{2}\delta m}\left[W_{k,\varepsilon,\varepsilon}^{\left(\tau,t\right)}F_{0k}-\frac{l_{k}^{4}\delta^{2}\beta_{k}^{2}m^{2}}{\hbar^{2}\left(\tau t\right)^{2\delta\varepsilon+\frac{7+\alpha}{\alpha}}}W_{k,\varepsilon-1,\varepsilon-1}^{\left(t,\tau\right)}G_{0k}\right],\nonumber \\
g_{k}= & \frac{\pi\beta_{k}\tau^{\delta}}{4}\left[\frac{\tau^{\delta\left(\varepsilon-1\right)}}{t^{\delta\left(\varepsilon-1\right)}}W_{k,\varepsilon,\varepsilon-1}^{\left(\tau,t\right)}F_{0k}-\frac{t^{\delta\varepsilon}}{\tau^{\delta\varepsilon}}W_{k,\varepsilon,\varepsilon-1}^{\left(t,\tau\right)}G_{0k}\right]\nonumber \\
- & \frac{i\pi}{4}\frac{\hbar\left(\tau t\right)^{\delta\varepsilon+\frac{3+\alpha}{\alpha}}}{l_{k}^{2}\delta m}\left[W_{k,\varepsilon,\varepsilon}^{\left(\tau,t\right)}F_{0k}+\frac{l_{k}^{4}\delta^{2}\beta_{k}^{2}m^{2}}{\hbar^{2}\left(\tau t\right)^{2\delta\varepsilon+\frac{7+\alpha}{\alpha}}}W_{k,\varepsilon-1,\varepsilon-1}^{\left(t,\tau\right)}G_{0k}\right],\label{B11}
\end{align}
where 
\begin{equation}
f_{0k}=f_{k}\left(\tau\right)=\frac{F_{0k}+G_{0k}}{2},\text{ \ }g_{0k}=g_{k}\left(\tau\right)=\frac{F_{0k}-G_{0k}}{2}.\label{B12}
\end{equation}


\begin{thebibliography}{10}
\bibitem{Peebles2003}P. J. E. Peebles and B. Ratra, Rev. Mod. Phys.\textbf{
75}, 559-606 (2003).

\bibitem{Oks2021}E. Oks, New Astron. Rev. \textbf{93}, 101632 (2021).

\bibitem{Riess1998}A. G. Riess \textit{et al.} {[}Supernova Search
Team{]}, Astron. J. \textbf{116}, 1009-1038 (1998).

\bibitem{Perlmutter1999}S. Perlmutter \textsl{et al.} {[}Supernova
Cosmology Project{]}, Astrophys. J. \textbf{517}, 565-586 (1999).

\bibitem{Ellis1999}G. F. R. Ellis and H. van Elst, NATO Sci. Ser.
C \textbf{541}, 1-116 (1999).

\bibitem{Leon1988}J. Ponce De Leon, Gen. Rel. Grav. \textbf{20},
539-550 (1988).

\bibitem{Wesson1994}P. S. Wesson, Astrophys. J. \textbf{436}, 547-550
(1994).

\bibitem{Wesson2002}S. S. Seahra and P. S. Wesson, Class. Quant.
Grav. \textbf{19}, 1139-1155 (2002).

\bibitem{Bellini2003}M. Bellini, Nucl. Phys. B \textbf{660}, 389-400
(2003) {[}erratum: Nucl. Phys. B \textbf{671}, 501-501 (2003){]}.

\bibitem{Parker1968}L. Parker, Phys. Rev. Lett. \textbf{21}, 562-564
(1968).

\bibitem{Parker1969}L. Parker, Phys. Rev. \textbf{183}, 1057-1068
(1969).

\bibitem{Parker1971}L. Parker, Phys. Rev. D \textbf{3}, 346-356 (1971)
{[}erratum: Phys. Rev. D \textbf{3}, 2546-2546 (1971){]}.

\bibitem{Grishchuk1989}L. P. Grishchuk and Y. V. Sidorov, Class.
Quant. Grav. \textbf{6}, L161-L165 (1989).

\bibitem{Grishchuk1990}L. P. Grishchuk and Y. V. Sidorov, Phys. Rev.
D \textbf{42}, 3413-3421 (1990).

\bibitem{Hu1994}B. L. Hu, G. Kang and A. Matacz, Int. J. Mod. Phys.
A \textbf{9}, 991-1008 (1994).

\bibitem{Ford2021}L. H. Ford, Rept. Prog. Phys. \textbf{84}, no.11,
116901 (2021).

\bibitem{Parker2009}L. E. Parker and D. Toms, \textit{Quantum Field
Theory in Curved Spacetime: Quantized Field and Gravity} (Cambridge
University Press, 2009).

\bibitem{DeWitt1975}B. S. DeWitt, Phys. Rept. \textbf{19}, 295-357
(1975).

\bibitem{Pedrosa2021}I. A. Pedrosa, B. F. Ramos and K. Bakke, Eur.
Phys. J. C \textbf{81}, no.8, 703 (2021).

\bibitem{Bertoni1998}C. Bertoni, F. Finelli and G. Venturi, Phys.
Lett. A \textbf{237}, 331-336 (1998).

\bibitem{Pedrosa2004}A. M. d. M. Carvalho, C. Furtado and I. A. Pedrosa,
Phys. Rev. D \textbf{70}, 123523 (2004).

\bibitem{Suresh2004}K. K. Venkataratnam and P. K. Suresh, Int. J.
Mod. Phys. D \textbf{13}, 239-252 (2004).

\bibitem{Suresh2004.2}P. K. Suresh, Int. J. Theor. Phys. \textbf{43},
425-436 (2004).

\bibitem{Alencar2012}G. Alencar, I. Guedes, R. R. Landim and R. N.
Costa Filho, EPL \textbf{98}, no.1, 11001 (2012).

\bibitem{Lewis1969}H. R. Lewis and W. B. Riesenfeld, J. Math. Phys.
\textbf{10}, 1458-1473 (1969).

\bibitem{Dodonov1975}V. V. Dodonov, I. A. Malkin and V. I. Man'ko,
Int. J. Theor. Phys. \textbf{14}, 37 (1975).

\bibitem{Dodonov2003}V. V. Dodonov and V. I. Man'ko (Ed.). \textit{Theory
of nonclassical states of light} (CRC Press, 2003).

\bibitem{Pereira2021}A. S. Pereira and A. S. Lemos, Phys. Lett. A
\textbf{405}, 127428 (2021).

\bibitem{Pereira2023}A. S. Pereira, A. S. Lemos and F. A. Brito,
Eur. Phys. J. Plus \textbf{138}, no.4, 363 (2023).

\bibitem{Baumann2009}D. Baumann, ``TASI Lectures on Inflation,''
arXiv:0907.5424 {[}hep-th{]}.

\bibitem{Andrews1999}M. Andrews, American Journal of Physics \textbf{67},
336-343 (1999).

\bibitem{Pereira2022}A. S. Pereira, A. S. Lemos and F. A. Brito,
Eur. Phys. J. Plus \textbf{137}, no.8, 957 (2022).

\bibitem{Dodonov1980}V. V. Dodonov, E. V. Kurmyshev and V. I. Man\textquotesingle ko, Phys. Lett. A \textbf{79} A, 150 (1980).

\bibitem{Fukui2001}T. Fukui, S. S. Seahra and P. S. Wesson, J. Math. Phys. \textbf{42}, 5195-5201 (2001).

\bibitem{Wesson1992}P. S. Wesson, Astrophys. J. \textbf{394}, 19 (1992).

\bibitem{Wesson1995}P. S. Wesson and H. Y. Liu, Astrophys. J. \textbf{440}, 1-4 (1995).

\bibitem{Clifton2012}T. Clifton, P. G. Ferreira, A. Padilla and C.
Skordis, Phys. Rept. \textbf{513}, 1-189 (2012).

\bibitem{Barcelo2005}C. Barcelo, S. Liberati and M. Visser, Living
Rev. Rel. \textbf{8}, 12 (2005).

\end{thebibliography}
\end{document}